\documentclass[aps,onecolumn,groupedaddress,nofootinbib,superscriptaddress,floatfix, reprint]{revtex4-1}

\RequirePackage[T1]{fontenc}

\usepackage{amsfonts}
\usepackage{amssymb}
\usepackage{amsmath}
\usepackage{graphicx}
\usepackage{flushend}
\usepackage{verbatim}

\usepackage[colorlinks,citecolor=blue,urlcolor=blue,linkcolor=blue]{hyperref}
\usepackage{float}
\usepackage{subcaption}

\newcommand{\BigO}[1]{\ensuremath{\operatorname{\mathcal{O}}\left(#1\right)}}

\begin{document}
	
\title{Stability of domain walls in models with asymmetric potentials}
	
\author{Tomasz Krajewski}
\email{Tomasz.Krajewski@fuw.edu.pl}
\author{Jan Henryk Kwapisz}
\email{Jan.Kwapisz@fuw.edu.pl}
\author{Zygmunt Lalak}
\email{Zygmunt.Lalak@fuw.edu.pl}
\author{Marek Lewicki}
\email{Marek.Lewicki@fuw.edu.pl}
	
\affiliation{Institute of Theoretical Physics, Faculty of Physics, University of Warsaw, ul. Pasteura 5, Warsaw, Poland}
	
\date{\today}
	
\begin{abstract}
We study the evolution of cosmological domain walls in models with asymmetric potentials. Our research goes beyond the standard case of spontaneous breaking of an approximate symmetry. When the symmetry is explicitly broken the potential exhibits nearly degenerate minima which can lead to creation of a metastable network of domain walls. The time after which the network will decay depends on the difference of values of the potential in minima, its asymmetry around the maximum separating minima and the bias of the initial distribution of the field. Effect of asymmetry around the maximum separating minima is novel one that we study with a new type of potential. Using numerical lattice simulations we determine relative importance of these factors on decay time of networks for generic potentials. We find that even very small departures from the symmetric initial distribution case lead to rapid decay of the domain wall network. As a result creation of a long lasting network capable of producing observable gravitational wave signals is much more difficult than previously thought. On the other hand details of the shape of the potential turn out to be much less important than was expected and the evolution of network from symmetric distribution is controlled by the difference of values of the potential in the minima.
\keywords{domain walls, topological defects, gravitational waves}
\end{abstract}

\maketitle

\section{Introduction}
Domain walls are topological defects\footnote{No continuous mapping exists that transforms single planar domain wall into trivial homogeneous configuration of the field.} which could be formed in the early Universe at boundaries of regions, called domains, in which certain field $\phi$ takes different vacuum expectation values. Usually domain walls are associated with spontaneous breaking of a discrete symmetry. In this case domains are patches of the Universe occupied by the field strength corresponding to different minima of the potential and domain walls are transition regions at which field strength smoothly interpolate between these minima.

Domain walls are usually considered to form during cosmological evolution from primordial fluctuations generated by some random process in the early Universe for example from quantum fluctuations during inflationary era. Domain walls form when fluctuations at the characteristic scale, corresponding to width of domain walls, cross the horizon. Percolation theory predicts that domain walls produced by a~stochastic process can form networks of twofold topologies:
\begin{itemize}
	\item separated bubbles of one vacuum submerged in the background of the another one,
	\item infinite domain walls stretching through the whole Universe,
\end{itemize}
depending on the initial contributions of the vacua \cite{PhysRevD.39.1558, Lalak:1994qt, Coulson:1995uq, Coulson:1995nv, Larsson:1996sp, Lalak:2007rs, Correia:2014kqa, Correia:2018tty}. During further evolution domains are stretched by the expansion of the Universe, thus both the surface and the curvature radius of domain walls grow. On the other hand, if a domain is too small and the gain of the energy contained in the enclosing wall from shrinking it overcomes the expansion, the domain will collapse.

Evolution of domain walls in the simplest case of degenerate minima and symmetric initial distribution leads to the so called scaling regime. This type of evolution is marked by simple scaling of average quantities such as surface area or averaged energy of domain walls per Hubble horizon. In the scaling regime number of domain walls in the horizon stays nearly constant, while average domain size and curvature radius are of order of Hubble horizon~\cite{Hindmarsh:1996xv}. Maintaining the scaling requires domain walls to frequently interact with each other, changing their configuration or collapsing into closed walls, to reduce their energy. The effective equation of state of a~network of cosmological domain walls is generically predicted with barotropic parameters $-2/3 < w_{DW} < -1/3$ \cite{Kolb:1990vq, Friedland:2002qs}. The energy density of the network of stable domain walls decreases (with the expansion) slower than the energy density of both: the radiation and the dust, so long lived domain walls tend to dominate the energy density of the Universe. Moreover the effective average pressure generated by the network is negative, thus it acts as Dark Energy. However, Dark Energy with such equation of state is ruled out by the present experimental data~\cite{PhysRevD.68.043509, Spergel:2003cb, Tegmark:2003ud, Conversi:2004pi, Avelino:2017dxn}. Moreover, domain walls which pose a significant fraction of the total energy density of the Universe at recombination would produce unacceptably large fluctuations of the Cosmic Microwave Background Radiation (CMBR).

However, if minima of the potential are not degenerate, pressure coming from difference of vacuum energy in different domains will act on domain walls and render them unstable\footnote{A~model with potential with non-degenerate minima does not provide a~solution of the equation of motion in Minkowski background which is a soliton and can be interpreted as a~domain wall. However, when the minima are nearly degenerate one can expect existence of a~solution which is slowly varying in time and similar to domain wall. Some authors distinguish these two cases and call the later solution a~domain wall like structure. We are considering cosmological evolution in non-trivial time dependent Friedman-Robertson-Walker background and we will use name domain walls in both cases.}. As a result they will annihilate on a~time scale which depends on the fraction of space occupied by the field strength corresponding to the global minimum of the potential, the width of the initial distribution of the field, the bias between the minima and the steepness of the potential on both sides of the local maximum separating the minima. While the effects of the bias in initial distribution and in the minima have been extensively studied (see \cite{PhysRevD.39.1558, Lalak:1994qt, Coulson:1995uq, Coulson:1995nv, Larsson:1996sp, Lalak:2007rs, PhysRevD.79.085007, Correia:2014kqa, Correia:2018tty}) the other two factors, haven't been discussed in the literature.

To study those we implement new type of potential that extends the analysis done before. The introduced family of potentials allow for independent parametric control of all four factors in our numerical simulations. In this work we study the impact of these factors on decay time. We discuss experimental prospects for detection through gravitational waves (GWs) signals produced upon domain walls annihilation. We find a hierarchy of strength of the dependence of the lifetime of the network on the parameters of interest. Every non-negligible departure from the exactly symmetric initial distribution case leads to rapid decay of the domain wall network, thus confirming the previous analytical and numerical studies \cite{PhysRevD.39.1558, Lalak:1994qt, Coulson:1995uq, Coulson:1995nv, Larsson:1996sp, Lalak:2007rs,Avelino:2008qy, Correia:2014kqa, Correia:2018tty}. On the other hand the asymmetry of the potential around maximum which has not been studied in the past turn out to introduce negligible effect on the stability of networks. Due to the described hierarchy compensating the influence of one of the factors by the another requires severe fine-tuning. As a result, formation of a long lasting network capable of producing an observable GW signal seems to be much more difficult than previously thought.

\section{Models of interest}
Domain walls in models with symmetric or nearly symmetric potentials were a main object of studies in the past. The evolution of these defects in models with asymmetric potentials is much less known although such models are frequently considered in modern cosmology. Asymmetric potentials with number of non-equivalent minima appear in many cosmological models at various energy scales. Let us present just a few examples of such models.

\subsection{Radiatively generated minima}
Even when tree level potential of the model has one minimum, radiative quantum corrections can result in formation of a~second minimum. This is exactly the case of the Standard Model (SM), where quantum corrections (mainly due to interaction with top quark) force running quartic coupling constant of the Higgs field to acquire negative values for renormalization scale of the order of $10^{10}\, \textrm{GeV}$. This leads to formation of the local maximum of the effective potential for the field strength of the order of zero of the running coupling (the position of the maximum is gauge dependent) and second deeper minimum at superplanckian field strengths. The dynamics of domain walls driven by this highly asymmetric potential was studied in details in \cite{Krajewski:2016vbr, Krajewski:2017czs, Krajewski:2019vix}. It is worth stressing that even though Higgs field breaks the symmetry, it is a gauge symmetry, thus all the minima of the effective potential connected by the symmetry transformation are physically equivalent. However, the radiative corrections leads to formation of a~second physically non-equivalent family of minima. In this case families of minima are not connected by any symmetry transformation and nothing forces them to be degenerate or the potential to be symmetric in between of them. Studying Higgs domain walls we have found that their dynamics differ from previously investigated models. This observation motivated us to a more detailed investigation of the influence of the asymmetry of the potential on the evolution of domain walls whose result we present in this manuscript.

\subsection{Run-away potentials}
\begin{figure}[!ht]
	\centering
	\includegraphics[width=3.375in]{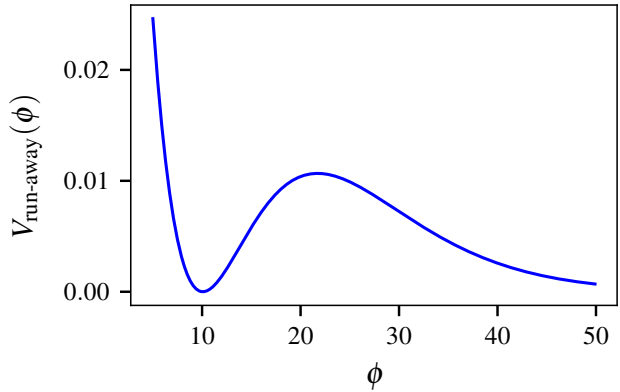} 
	\caption{Shape of the potential used in \cite{Lalak:2007rs} for simulating the dynamics of domain walls in models with run-away potentials.\label{fig:run-away_potential}}
\end{figure}

Run-away potentials frequently appear in models of dynamical supersymmetry breaking and play an important role in modern attempts to non-perturbative supersymmetry breaking and moduli stabilization. The local minima in such potentials are separated by a~barrier from otherwise monotonically decreasing yet bounded from below potential. It has been pointed out by Dine \cite{Dine:1982ah} that spatial inhomogeneities may help to stabilize the moduli at shallow but finite minima. The dynamics of domain wall like structures in the model with the runaway potential of the form
\begin{multline}
V_{\text{run-away}}(\phi) = \frac{1}{2\phi} \left(A(2\phi + N_1) e^{-\phi/N_1} \right. \\ \left.- B (2\phi + N_2) e^{-\phi/N_2} \right), \label{eq:run-away_potential}
\end{multline}
expected in a wide class of supersymmetry breaking models based on gaugino condensation was investigated in \cite{Lalak:2007rs}. The shape of this potential for the choice of parameters $A$, $B$, $N1$ and $N2$ used in \cite{Lalak:2007rs} is presented in the Fig~\ref{fig:run-away_potential}. It has been found that in this scenario evolution of the network of inhomogeneities is very similar to a better known case of symmetric potentials. However, dependence on parameters of this model was not studied due to limited available computational resources. Thus, it cannot be determined if this is a generic feature of this class of models.

\subsection{Models of out-of-equilibrium phase transitions}
In a first-order phase transition a~metastable vacuum with expectation value of the field strength equal to a~local minimum of a~potential is separated from the so called true vacuum corresponding to the global minimum of the potential by a potential barrier. Initially the field is assumed to be trapped in false vacuum, due to its earlier evolution. Thus, time depended potentials are considered, usually realized by temperature dependence. In the early Universe, when the temperature was high, thermal corrections to the effective potential modified it in such a way that it had only one minimum. The field strength evolved toward this high temperature minimum. During evolution of the Universe, when the temperature was decreasing, second minimum was formed. If the potential barrier developed before the second minimum became the global one, the field was trapped in metastable vacuum, due to inability of classically traversing the barrier. As a result the unstable vacuum decays through nucleation of bubbles, corresponding to the field trapped in the false vacuum quantum mechanically tunneling through the barrier \cite{PhysRevD.16.1248, Callan:1977pt, Linde:1981zj}.

After nucleation, bubbles grow until they collide, eventually converting the whole Hubble volume into the new phase. During collision phase of the transition when many bubbles collided with each other, an~intermediate state similar to a~network of cosmological domain walls is formed. The main difference between this state and the network formed from superhorizon fluctuations is that boundaries of domains keep large velocities generated during expansion of bubbles before collisions.

\subsection{Axion monodromy models}
\begin{figure}[!ht]
	\centering
	\includegraphics[width=3.375in]{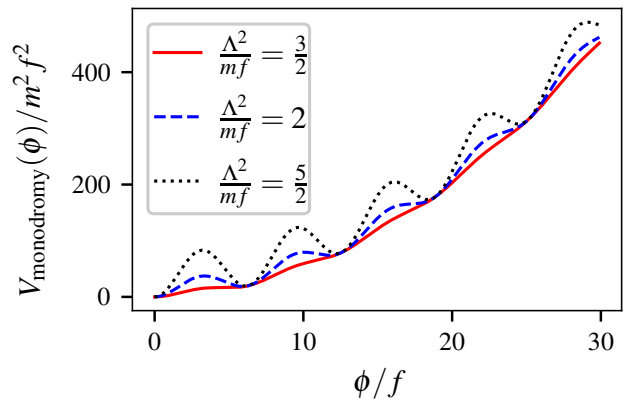}
	\caption{Shape of axion monodromy type potentials given by eq. \eqref{eq:monodromy_potential} for various value of the fraction $\frac{\Lambda^2}{m f}$.\label{fig:monodromy_potential}}
\end{figure}

Axions have a~periodic potential conventionally parameterized as cosine-type:
\begin{displaymath}
V_{\text{axion}}(\phi) = \Lambda^4 \left[1- \cos\left(\frac{\phi}{f}\right) \right],
\end{displaymath}
where $f$ is the decay constant and $\Lambda$ is the scale of nonperturbative effects that generate the potential. Such a form of periodic potentials is derived in dilute instanton gas approximation.
The dynamically generated potential breaks shift symmetry of axion (a~pseudo Nambu-Goldstone boson of spontaneously broken global symmetry) to its discrete $\mathbb{Z}$ subgroup acting as $\mathbb{Z} \ni n \colon \phi \mapsto \phi + 2\pi f n$. Due to this symmetry the axion field strength is bounded to $0 \le \phi < 2 \pi f$.
In the presence of a monodromy the potential contains additional terms which explicitly break remaining discrete symmetry and enlarge field strength range. The monodromy term is usually chosen as a~monomial, frequently as the quadratic term $\frac{1}{2} m^2 \phi^2$, less often as the linear $g \phi$. Let us concentrate on the former choice which give the potential in the form:
\begin{equation}
V_{\text{monodromy}}(\phi) = m^2 \phi^2 +  \Lambda^4 \left[1- \cos\left(\frac{\phi}{f}\right) \right], \label{eq:monodromy_potential}
\end{equation}
For the proper choice of values of parameters of the model, the potential consist of quadratic potential decorated with wiggles coming from the periodic term as can be seen in the figure \ref{fig:monodromy_potential}. The potential develops a family of local minima separated by asymmetric potential barriers whose number depends on the relative strength of both terms quantified by the fraction $\frac{\Lambda^2}{m f}$.

Axion like particles can play important role in cosmology (for a review see e.g.\cite{Marsh:2015xka}). If produced in appropriate abundance, they are good candidates for cold dark matter \cite{Abbott:1982af, Preskill:1982cy, Dine:1982ah, Arias:2012az}. Moreover, axion potentials with monodromy are considered as suitable for inflationary models (also models with non-mininal couplings \cite{McDonough:2020gmn}). Presence of local minima which drift slowly rolling inflaton to ultra slow-roll regime leading to enhancement of primordial scalar fluctuations was proposed as a origin of primordial black holes \cite{Ballesteros:2017fsr, Ozsoy:2018flq, Mishra:2019pzq, Ballesteros:2019hus}.
The production of gravitational waves emitted when inflaton field traverse features of axion monodromy potential was studied in \cite{McAllister:2008hb, Ozsoy:2020ccy, Ozsoy:2020kat}. Moreover, possibility of population of global minimum of monodromy type potentials and neighbouring local ones during reheating due to oscillations of inflaton was discussed in \cite{Hebecker:2016vbl}.

On the other hand, the latter choice with the linear symmetry breaking term leads to the potential in the form:
\begin{equation}
V_{\text{relaxion}}(\phi) = g \phi +  \Lambda^4 \left[1- \cos\left(\frac{\phi}{f}\right) \right]. \label{eq_relaxion_potential}
\end{equation}
Potentials of this type play a crucial role in cosmological relaxation models proposed in order to solve hierarchy problem of the Standard Model Higgs boson mass \cite{Graham:2015cka, Patil:2015oxa, Gupta:2015uea, Espinosa:2016voy}. In this scenario the dimensionfull coupling $g$ is assumed to depend on the Higgs field strength. Then, the wiggles produced by the periodic term stop the slowly rolling field $\phi$ (called relaxion) at the point which generates proper Higgs boson vacuum expectation value. The original idea based on QCD axion was broadly extended and modified recently \cite{Hardy:2015laa, Fonseca:2018xzp, Ibe:2019udh, Fonseca:2019ypl, Fonseca:2019lmc}.

Topological defects were studied with various methods in the past. The Nambu-Goto effective action was first used in the case of cosmic strings~\cite{PhysRevD.54.2535} and later generalized to domain walls~\cite{Avelino:2005kn,Avelino:2005pe}. In this approach topological defects are treated as very thin, thus effects of finite thickness of these structures are neglected. This assumption is satisfied the better the larger are structures and distances between them compared to their thickness. Thus, this method is useful in investigation of the late evolution of networks of topological defects when they are diluted and stretched by the expansion of the Universe.

From the Nambu-Goto action the velocity-dependent one scale (VOS) can be derived \cite{Avelino:2005kn, Avelino:2008qy, Avelino:2010qf, Leite:2011sc, Martins:2016ois} under the assumption that only one scale is present in the problem i.e. the curvature radius, distances between defects and Hubble radius are of the same order. It is a semi-analytical model which needs to be calibrated by numerical methods. It correctly reproduces the scaling regime in which the number of defects in each Hubble radius stays nearly constant and the time at which it ends i.e. decay of the network for models with (nearly) symmetric potentials. Obviously, this semi-analytical method is not able to properly model subtle effects caused by the shape of the potential.

Due to non-linear character of topological defects, the most reliable methods of investigating their dynamics are lattice simulations. In these simulations the equation of motion of field forming defects is numerically integrated by the finite difference method. Majority of lattice numerical simulations of the dynamics of domain walls performed in the past \cite{Press:1989yh, Coulson:1995uq, Larsson:1996sp, Lalak:1996db, Oliveira:2004he, Lalak:2007rs, Kawasaki:2011vv, Leite:2011sc, Hiramatsu:2013qaa} were based on simple potentials known in analytical form. Most attention was given to domain walls in the case of spontaneous breaking of global discrete symmetries. In this scenario, minima of the potential of the model are degenerated. When the symmetry is weakly, explicitly broken the symmetric minima are nearly degenerated. Up to our knowledge, in all previous studies one symmetry breaking term was introduced into the potential whose coupling constant controlled both degeneracy on minima and a~shape of the potential in between them or an asymmetric potential motivated by certain model was studied. In order to understand what really determine the fate of network of domain walls --- is it a degeneracy of the minima or the shape of the potential around the top of the barrier or rather both --- we go beyond these simple models in our studies.

During the process of the decay of domain walls the energy of the field is transferred to other degrees of freedom, and a~part of it will be carried by gravitational waves (GWs). The recent observation of GWs at the LIGO and the Virgo experiments \cite{Abbott:2016blz} promoted spectrum of GWs to one of the most promising cosmological observable for many models. GWs can in principle probe domain walls in the early Universe. Moreover, GWs produced from networks of domain walls could partially polarise CMBR, marking it with a distinctive pattern. We try to estimate a~spectrum of GWs produced during the decay of domain walls using semi-analytical approximations introduced in previous studies \cite{Kitajima:2015nla, Hiramatsu:2013qaa}.

The paper is organized as follows. In section \ref{model} we introduce the analytic form of potentials which we used to model asymmetry of potentials through out the paper. The method we use to estimate the width of domain walls is presented in subsection \ref{width}. Set of asymmetric potentials that we used in our numerical simulations is given in subsection \ref{family}. Section \ref{decay_time} is dedicated to estimation of the lifetime of networks of domain walls. We discuss the dependence of the lifetime on initial conditions: the average value of the field strength and its standard deviation at the initialization and on scale of asymmetry of the potential of the model. Dependence of duration of scaling regime on parameters of the model and initial conditions is studied in section \ref{scaling_regime}. We discusses possible influence of asymmetry of the potential on the spectrum of GWs emitted from the network in section \ref{spectrum}. We conclude in section \ref{summary}. In appendix \ref{app:toy_model} we discuss the origin of quantity called width of domain walls in the case of simple, toy model in which analytic expression is known.
In appendix \ref{app:parameters} we present the parameters of the potential and initial conditions for the field that we have used in our simulations.

\section{Model of asymmetric potentials\label{model}}
In the past, mainly spontaneous breaking of $\mathbb{Z}_2$ symmetry in a~simple model defined by the Lagrangian density of the form:
\begin{equation}
\mathcal{L} = \frac{1}{2} \partial_\mu \phi \partial^\mu \phi - V_{\mathbb{Z}_2} := \frac{1}{2} \partial_\mu \phi \partial^\mu \phi - V_0 \left(\frac{\phi^2}{{\phi_0}^2}-1\right)^2. \label{symmetric_lagrangian_density}
\end{equation}
was studied in lattice simulations. $\mathbb{Z}_2$ symmetry guarantees that minima of the potential $V_{\mathbb{Z}_2}$ are degenerated and the potential is symmetric around the local maximum ($\phi = 0$).

In order to avoid experimental constraints the explicit symmetry breaking is usually considered. In the past studies the symmetry breaking term was added to the potential:
\begin{equation}
V_{EB} (\phi) =V_{\mathbb{Z}_2} (\phi) + \epsilon V_0 \phi. \label{toy_lagrangian_density}
\end{equation}
to destabilize domain walls. $\epsilon$ is a~parameter that determines the strength of explicit symmetry breaking. It was found \cite{Lalak:2007rs} that if the $\mathbb{Z}_2$ is explicitly broken, domain walls interpolating between minima of the potential are unstable and they annihilate on a~time scale which depends on the fraction of the space occupied by the field strength corresponding to the global minimum of the potential, the bias between minima (i.e. the difference between values of the potential at minima) and the value of the derivative on both sides of the local maximum separating the minima. However, the relation between influence of the last and the others factors was not determined so far. The model given by the eq. \eqref{toy_lagrangian_density} is unsuitable for such studies. Both the difference of values of the potential at the minima and the asymmetry of the potential around its local maximum are controlled by the value of one parameter $\epsilon$ and cannot be changed independently.

The aim of this paper is to overcome limitations of the simple model \eqref{toy_lagrangian_density} and prepare the set of potentials convenient for further studies in lattice simulations. From the point of view of lattice simulations it is convenient to define the potential of the model by its derivative which is directly used in simulations. The equation of motion for the symmetry breaking field with the canonical kinetic term and general potential $V$ is of the form:
\begin{equation}
\frac{\partial^2 \phi}{\partial \eta^2} + \frac{2}{a} \left(\frac{d a}{d \eta}\right) \frac{\partial \phi}{\partial \eta} - \Delta \phi + a^2 \frac{\partial V}{\partial \phi}=0, \label{eom}
\end{equation}
assuming the Friedman-Robertson-Walker metric background:
\begin{equation}
g = dt^2 - a^2(t) \delta_{ij} dx^i dx^j = a^2(\eta) \left(d\eta^2 -\delta_{ij} dx^i dx^j\right),
\end{equation}
where Latin indices correspond to spatial coordinates, $t$ is cosmic time and $\eta$ denotes conformal time (such that $d \eta = \frac{1}{a(t)} dt$). The equation \eqref{eom} which depends on the derivative of the potential $\frac{\partial V}{\partial \phi}$ is solved in lattice simulations using the finite difference scheme. Moreover, the position of the local extrema of the potential are easier to determine from its derivative.

Thus, we decided to give the potential by passing its derivative. We assumed the derivative in the form:
\begin{equation}
\frac{\partial V_{AS}}{\partial \phi} (\phi) := V_0 (\phi - a)(\phi - b)(\phi - c)\left(e^2 (\phi - d)^2 + 1\right),
\end{equation}
where $a$, $b$, $c$ determine positions of the extrema of the potential and parameters $e$, $d$ controls the shape of the potential. Then, the potential $V_{AS}$ takes the complicated form:
\begin{multline}
V_{AS}(\phi) = \frac{V_0}{60} \phi  \left(-60 a b c \left(d^2 e^2+1\right)\right. \\
+15 \phi ^3 \left(e^2 \left(2 d (a+b+c)+a b+a c+b c+d^2\right)+1\right)\\
-20 \phi ^2 \left(e^2 \left(d^2 (a+b+c)+2 d (a (b+c)+b c)+a b c\right) \right.\\
\left.+a+b+c\right)\\
+30 \phi  \left(d e^2 (a d (b+c)+2 a b c+b c d)+a b+a c+b c\right)\\
\left. -12 e^2 \phi ^4 (a+b+c+2 d)+10 e^2 \phi ^5 \right). \label{eq:asymmetric_potential}
\end{multline}
We quantitatively estimate the asymmetry of the potential around the local maximum as a~value of the third derivative of the potential:
\begin{multline}
\frac{\partial^3 V_{AS}}{\partial \phi^3}(\phi) = 2 V_0 \left(e^2 (a-\phi ) (\phi -b) (c+2 d-3 \phi )\right.\\
+(-a-b+2 \phi ) \left(e^2 (d-\phi ) (2 c+d-3 \phi )+1\right)\\
\left.+(\phi -c) \left(e^2 (d-\phi )^2+1\right)\right)
\end{multline}
at the maximum.

Our aim is to find a family of potentials of the form \eqref{eq:asymmetric_potential} with given difference of values at minima $V_{AS}(b) - V_{AS}(a)$ (where we assume without loss of generality that $b$ and $a$ are minima of the potential and $c$ is its local maximum) and the value of the third derivative at the maximum $\frac{\partial^3 V_{AS}}{\partial \phi^3}(c)$. However, these conditions are not sufficient to perform simulations whose results will reveal dependence of the dynamics of domain walls on the asymmetry of the potential. Above those, the same energy scale for all cases given by the width of the walls which we will define in the next section \ref{width} is needed to compare the results.

\subsection{The width of domain walls\label{width}}
The estimation of the physical width of domain walls is critical for numerical simulations of their dynamics. The width must be at least a~few times larger than the lattice spacing (i.e. the physical distance between neighbouring points) used in the simulation in order to assure sufficient accuracy to model profiles of walls. On the other hand, if we choose the lattice spacing too small (walls will spread over too many lattice points) only few walls will fit into the finite lattice. If only small number of walls will be present on the lattice, then dynamics of the network of domain walls will be reproduced poorly in the simulation. Many authors \cite{Press:1989yh,Coulson:1995uq,Lalak:1996db,Oliveira:2004he,Lalak:2007rs,Kawasaki:2011vv,Leite:2011sc,Hiramatsu:2013qaa} used simulations with the physical width of walls varying from 2 to 100 lattice spacing.

We estimate the width of domain walls using the approach presented in the appendix \ref{app:toy_model} based on the first integral of the equation of motion. Firstly, we calculate the value $\phi_2$ of the field which gives the same value of the potential as the value $\phi_{min}$ taken by the field in the local minimum and bigger than the local maximum. Next, we use integral expression (analog of the eq. \eqref{eq:tension}):
\begin{equation}
\Sigma(\varphi_1,\varphi_2):= \int_{\varphi_1}^{\varphi_2}\frac{V_{AS}(\varphi) d\varphi}{\sqrt{2\left(V_{AS}\left(\varphi\right)-V_{AS}(\phi_{min})\right)}}, \label{eq:generic_tension}
\end{equation}
to compute tension (surface energy density) of walls $\sigma_{wall} := \Sigma(\phi_{min}, \phi_2)$. Then, we use generalization of eq. \eqref{eq:soliton} for calculation of distance in the physical space in the direction perpendicular to the wall:
\begin{equation}
X(\varphi_2)-X(\varphi_1)= \int_{\varphi_1}^{\varphi_2} \frac{d\varphi}{\sqrt{2\left(V_{AS}\left(\varphi\right)-V_{AS}(\phi_{min})\right)}}.\label{eq:approx_solution}
\end{equation}
Finally, the width of walls is given as $w:= X(\tilde{\varphi_2})-X(\tilde{\varphi_1})$ for the pair of field strengths $\tilde{\varphi_1}$ and $\tilde{\varphi_2}$ such that $V_{AS}\left(\tilde{\varphi_1}\right)=V_{AS}\left(\tilde{\varphi_2}\right)$ and
\begin{equation}
\frac{\Sigma(\tilde{\varphi_1},\tilde{\varphi_2})}{\Sigma(\phi_{min},\phi_2)}\approx 97 \%.
\end{equation}
Thus, the width of walls is a~characteristic length of the distance in the direction perpendicular to the wall at which majority of potential energy density is stored.
Basing on the results of \cite{Lalak:2007rs} we chose the value of the width of domain walls to be $w = 5$.

\subsection{Family of asymmetric potentials\label{family}}
Our aim is to prepare a~family of potentials with given difference of the values at minima $\delta V$, the value of the third derivative at local maximum $d3V$ and the constant width equal to $5$. In order to resolve this problem we need to solve the following set of equations:
\begin{equation}
\begin{split}
\delta V &= V_{AS}(b) - V_{AS}(a),\\
d3V &= \frac{\partial^3 V}{\partial \phi^3} (c),\\
5 &= w, \label{eq:set}
\end{split}
\end{equation}
were the width is calculated numerically using the algorithm presented in subsection \ref{width}.

\begin{figure}[!ht]
	\centering
	\includegraphics[width=3.375in]{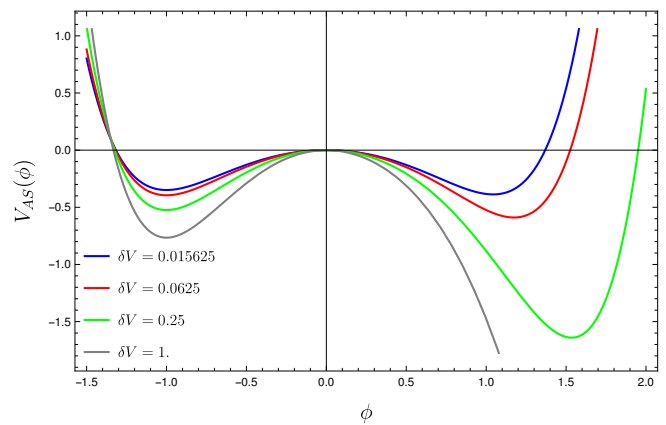}
	\caption{Shape of potentials obtained as solutions to the problem \protect\eqref{eq:final_set} for various values of $\delta V$ and $d3V=0$.\label{fig:deltaV_potentials}}
\end{figure}

\begin{figure}[!ht]
	\centering
	\includegraphics[width=3.375in]{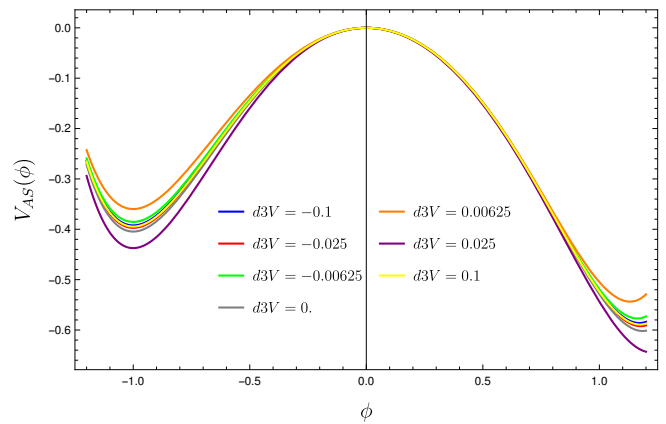}
	\caption{Shape of potentials obtained as solutions to the problem \protect\eqref{eq:final_set} for various values of $d3V$ and $\delta V = 0.0625$.\label{fig:d3V_potentials}}
\end{figure}

However, presented set of equation \eqref{eq:set} is not fully determined due to certain symmetries. First of all, the potential has translational symmetry in $\phi$ value which we fix by assuming that the local maximum lies at the value $c=0$. Secondly the field strength may be rescaled $\phi \mapsto \alpha \phi$. This rescaling combined with $V_0 \mapsto \alpha^{-4} V_0$ and $e \mapsto \alpha^{-1} e$ lefts form of the set of equations unchanged. In order to get rid of this symmetry we take the one minimum to be at $a=-1$. Finally, the dynamics of domain walls will stay unchanged if we add a constant value to the potential, as far as one neglects back-reaction from gravity. Thus, we assume that $V_{AS}(0)=0$. In addition we set $e=1$. With these assumptions the considered equations take simplified form:
\begin{equation}
\begin{split}
\delta V =& -\frac{V_0}{60} (b+1)^3 \left(5 (b-1) d^2+2 (4-3 b) b d \right.\\
&\left.+b (b (2 b-3)+8)-6 d-7\right),\\
d3V =& 2 V_0 \left(-b (d-1)^2+d^2+1\right),\\
5 =& w.\label{eq:final_set}
\end{split}
\end{equation}
By dividing the first equation by the second, one obtains the equation that is independent of $V_0$, connects $b$ and $d$ and can be solved in favor of $d$. Moreover it can be easily shown that the width scales as $w \propto V_0^{-1/2}$ with $V_0$. Finally, one need to solve numerically two equations from which one is given by the complicated numerical algorithm. Examples of potentials obtained in our procedure are plotted in figures \ref{fig:deltaV_potentials} and \ref{fig:d3V_potentials}. The figure \ref{fig:deltaV_potentials} presents solutions with various differences of values of potentials in their minima and with third derivative vanishing at the local maximum. The figure \ref{fig:d3V_potentials} shows solutions for $\delta V = 0.0625$ and various values of the third derivative.

The values of parameters $b$, $d$ and $V_0$ obtained by described procedure and used in our numerical lattice simulations are presented in appendix \ref{app:parameters}. Our final results are presented in the table \ref{tb:parameters} containing numerically obtained solutions for $b$, $d$ and $V_0$ for various values of $d3V$ and $\delta V$.

\section{Influence of asymmetry of a potential on decay time of the network\label{decay_time}}

Cosmological domain walls are subject to many experimental constraints. Generally, the energy density of a network of domain walls is predicted to decreases slower than the energy density of both: the radiation and the dust, so long lived domain walls would dominate the Universe. The equation of state of the network of domain walls is restricted to $-2/3 < w_{DW} < -1/3$, which is ruled out by the present data for a~single component Dark Energy. Domain walls which lived long enough to be present during the recombination would produce unacceptably large fluctuations of CMBR. This results in the Zel'dovich bound on the characteristic scale of physics producing domain walls present during recombination of the order of $<1$ MeV \cite{Zeldovich:1974uw, Lazanu:2015fua}.

We concentrated on the evolution of domain walls during radiation domination era which are less constrained experimentally. In our numerical simulation based on PRS algorithm \cite{Press:1989yh} we assumed that
\begin{equation}
a(\eta) = \frac{\eta}{\eta_{start}}, \label{eq:scale_factor}
\end{equation}
according to the fact that scale factor $a$ scales as $a \propto \eta$ with conformal time $\eta$ during radiation domination.

On the basis of discussion in \cite{Coulson:1995uq} we assumed that an~initial distribution of the field strength is given by the Gaussian probability distribution
\begin{equation}
P(\phi) = \frac{1}{\sqrt{2 \pi} \sigma} e^{-\frac{\left(\phi -\theta\right)^2}{2 \sigma^2}}. \label{eq:gauss_distribution}
\end{equation}
We studied the evolution of networks of domain walls initialized with different values of $\theta$ and $\sigma$ in order to accommodate variety of processes leading to formation of walls. According to \cite{Lalak:2007rs} the final state and length of decay time of networks of domain walls depend on the fraction of the space occupied by the field strength corresponding to the basin of attraction of the global minimum of the potential, the bias between minima (i.e. the difference between values of the potential at minima) and the value of the derivative on both sides of the local maximum separating the minima.

Our simulations were started with three initial conditions:
\begin{itemize}
	\item initial conformal time $\eta_{start}$,
	\item initial mean value of the field strength $\theta$,
	\item initial standard deviation $\sigma$.
\end{itemize}
Initial conformal time $\eta_{start}$ is determined by the time at which domain walls are formed in the early Universe. However, the initial time of the simulation must be earlier, in order to smooth out the initial numerical fluctuations by the field evolution. The time of the formation of a network of domain walls can be determined from the evolution of statistical quantities calculated in the simulation. Our simulations were run with the initial conformal time equal to $\eta_{start} = c\, l$ where $l$ is the lattice spacing.

Initial conditions cannot be deduced from the dynamics of domain walls by itself and must be derived from a~model of the evolution of the early Universe (for example an~inflationary model). Hence, our results can also be thought as a constraint on the space of models of the early Universe.

For each set of initialization conditions we run five simulation on the lattice of the size of $512^3$ if decay time of the network is longer than $256\; c\, l$ and only one simulation on the lattice of the size of $512^3$ and four on smaller lattice of the size $256^3$ otherwise. This choice is motivated by the fact that conservatively the dynamic range of lattice simulations with periodic boundary conditions is bounded and conformal time need to be smaller than size of the lattice (multiplied by the speed of the light). On the other hand we cannot guarantee that decay times longer than $512$ are reliably computed and networks will not decay later than observed in simulations. Fortunately, only a very small fraction of simulated cases are touched by this issue. Described five runs were the base for analysis of statistical fluctuations of results obtained from simulations which proofed that their are highly consistent.

\begin{figure*}[!ht]
	\begin{subfigure}[t]{0.5\textwidth}
		\label{fig:deltaV_d3V_init_mean=0,0_init_sigma=1,0}
		\includegraphics[width=3.375in]{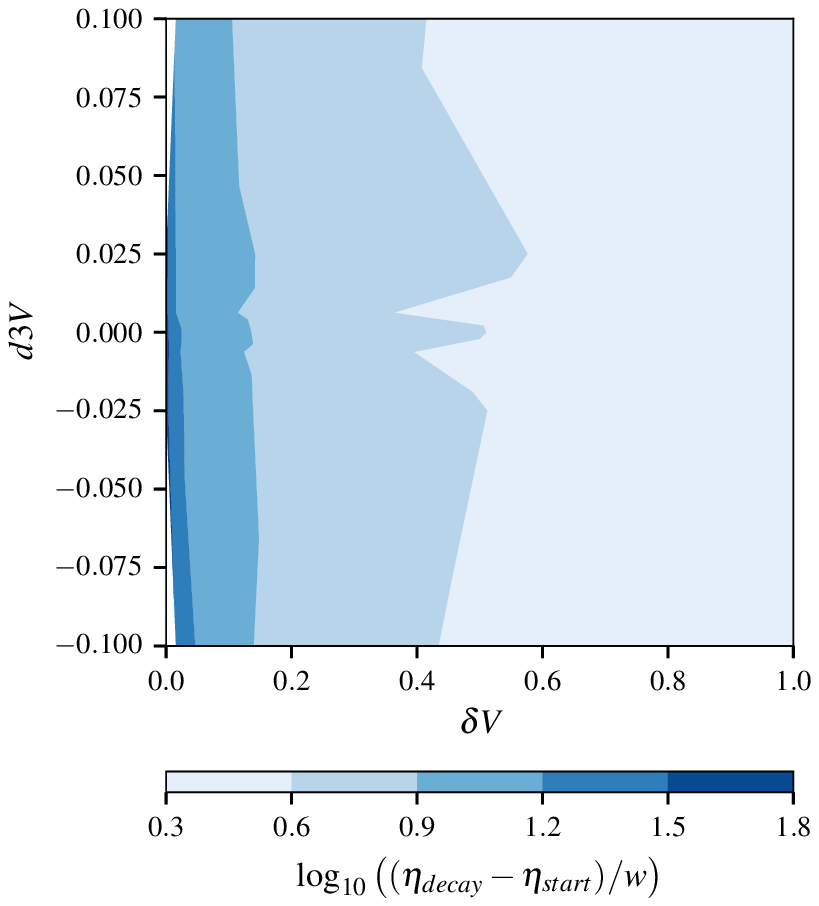}
	\end{subfigure}\hfill%
	\begin{subfigure}[t]{0.5\textwidth}
		\label{fig:deltaV_d3V_init_mean=0,0_init_sigma=0,25}
		\includegraphics[width=3.375in]{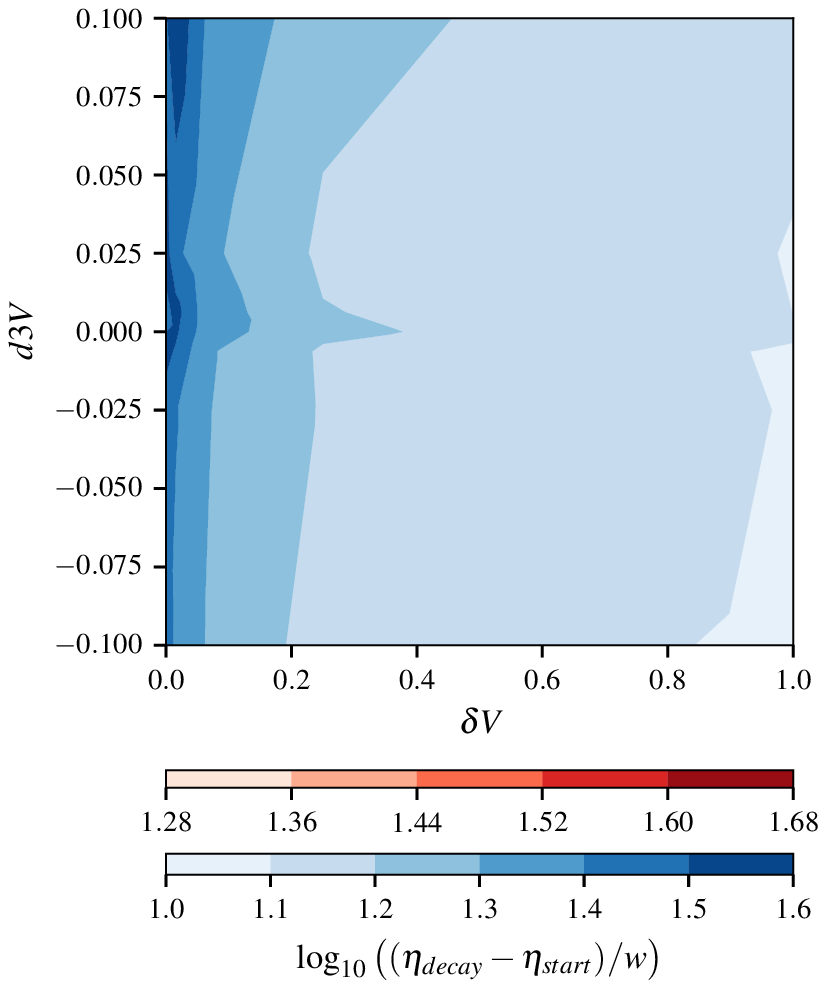}
	\end{subfigure}\hfill \\
	\begin{subfigure}[t]{0.5\textwidth}
		\label{fig:deltaV_d3V_init_mean=0,0_init_sigma=0,125}
		\includegraphics[width=3.375in]{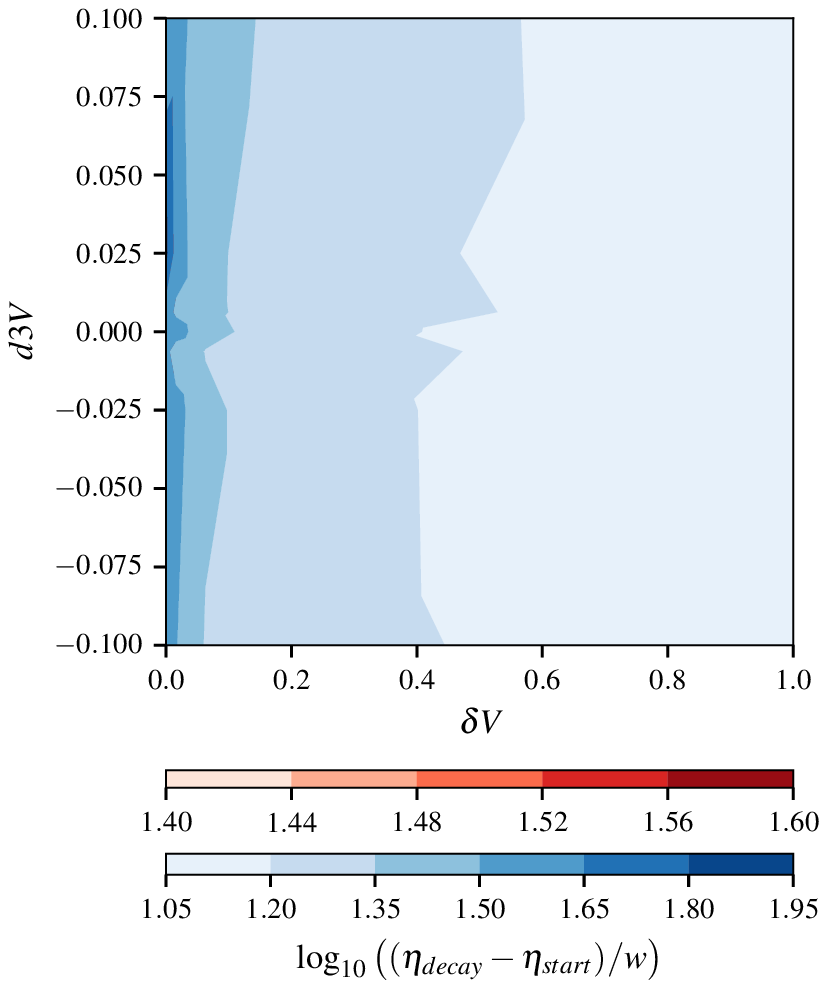}
	\end{subfigure}\hfill%
	\begin{subfigure}[t]{0.5\textwidth}
		\label{fig:deltaV_d3V_init_mean=0,0_init_sigma=0,0625}
		\includegraphics[width=3.375in]{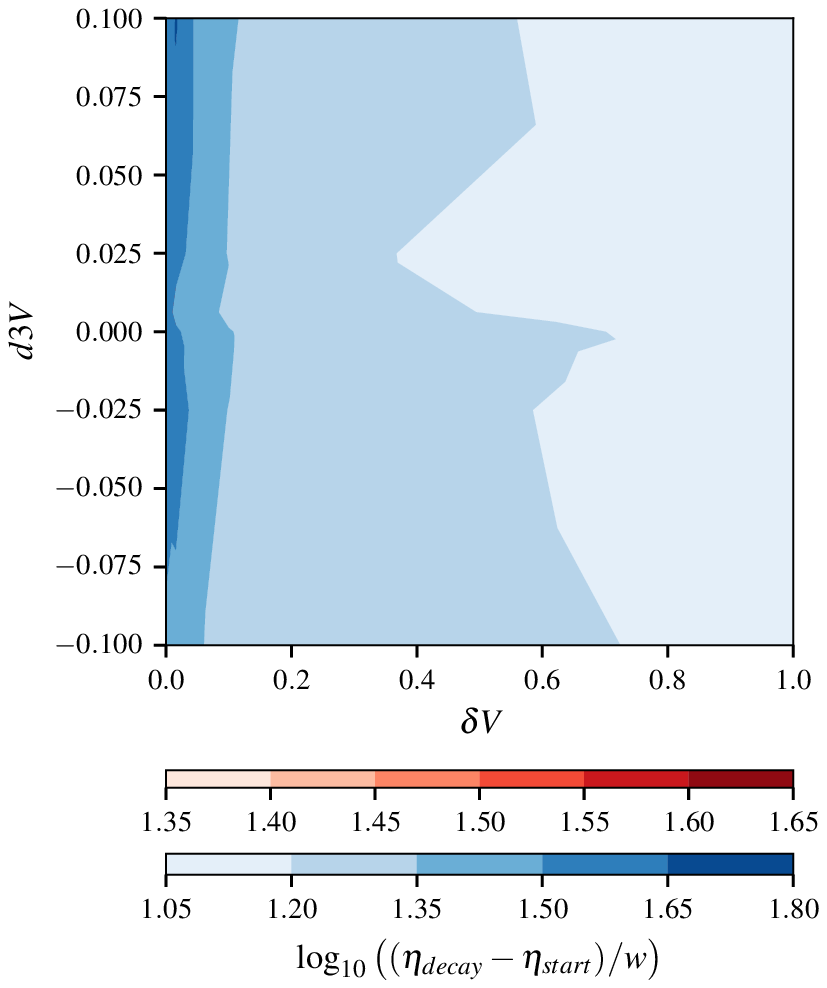}
	\end{subfigure}\hfill%
	\caption{Dependence of the decay time $\eta_{dec} - \eta_{start}$ on $\delta V$ and $d3V$ for initialization distribution with $\theta = 0$, $\sigma = 1$ (top left), $\sigma = 0.25$ (top right), $\sigma = 0.125$ (bottom left) and $\sigma = 0.0625$ (bottom right). \protect\label{fig:decay_time_deltaV_d3V}}
\end{figure*}

In the figure \ref{fig:decay_time_deltaV_d3V} we presented length of decay time of networks as a~function of parameters $\delta V$ and $d3V$ for unbiased initial distributions with four different values of the standard deviation $\sigma = 1$, $\sigma = 0.25$, $\sigma = 0.125$ and $\sigma = 0.0625$. Blue regions in these plots were extrapolated from simulations in which evolution of networks ended in the basin of the attraction of the global minimum of the potential and red ones from networks decaying to the local minimum. It can be deduced from plots of \ref{fig:decay_time_deltaV_d3V} that the main factor determining lifetime of networks is the difference of values of the potential in its minima. Much smaller effect, however still recognizable, is associated with asymmetry of potential around the local maximum parameterized in this case by $d3V$.

For wide initial distributions ($\sigma = 1$) networks tend to decay to global minimum of the potential and only networks in models with $\delta V = 0$ decayed into unstable vacuum. These observations are in agreement with naive expectation that for wide distributions evolution of the network will probe the shape of the potential at large distance, especially around minima more efficiently, thus the shape in neighborhood of the maximum will not took much effect. Decay times are slightly longer for positive $d3V$, thus for models in which potential is steeper on the side of the local minimum. This is again consistent with naive prediction that steepness of the potential around the maximum opposite in the direction to the evolution of the network may slow down the process of the decay. With decreasing width of the initial distribution this effect increases. Moreover, the range of parameter $\delta V$ for which networks decay to the unstable vacuum increases.

\begin{figure*}[!ht]
	\begin{subfigure}[t]{0.5\textwidth}
		\label{fig:deltaV_init_mean_d3V=0,0_init_sigma=1,0}
		\includegraphics[width=3.375in]{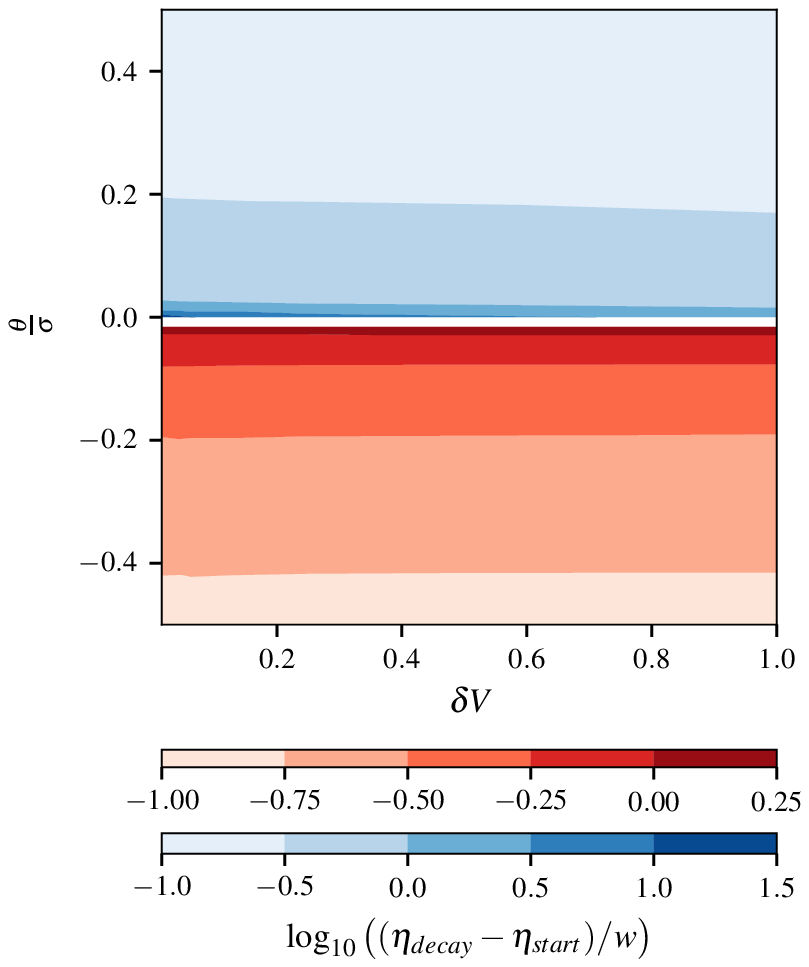}
	\end{subfigure}\hfill%
	\begin{subfigure}[t]{0.5\textwidth}
		\label{fig:deltaV_init_mean_d3V=0,0_init_sigma=0,0625}
		\includegraphics[width=3.375in]{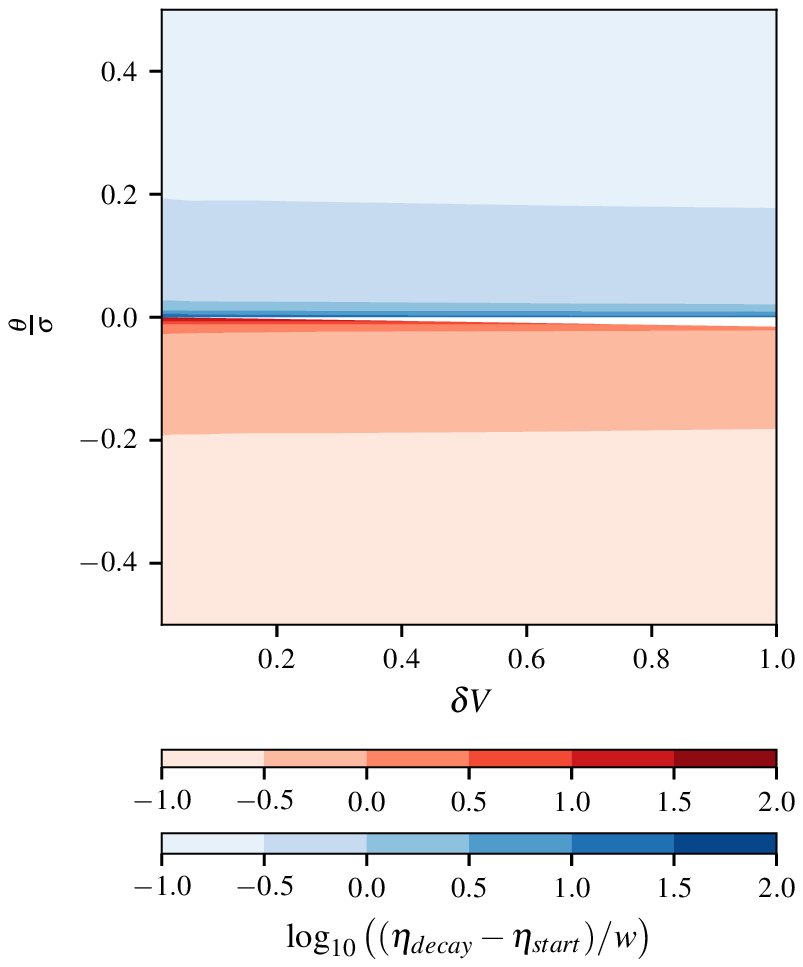}
	\end{subfigure}\hfill%
	\caption{Dependence of the decay time $\eta_{decay} - \eta_{start}$ on $\delta V$ and the mean value of initialization distribution $\theta$ with $\sigma = 1$ (left panel) and $\sigma = 0.0625$ (right panel) for models with $d3V = 0$. \protect\label{fig:decay_time_deltaV_init_mean}}
\end{figure*}

\begin{figure*}[!ht]
	\begin{subfigure}[t]{0.5\textwidth}
		\label{fig:d3V_init_mean_deltaV=0,0_init_sigma=1,}
		\includegraphics[width=3.375in]{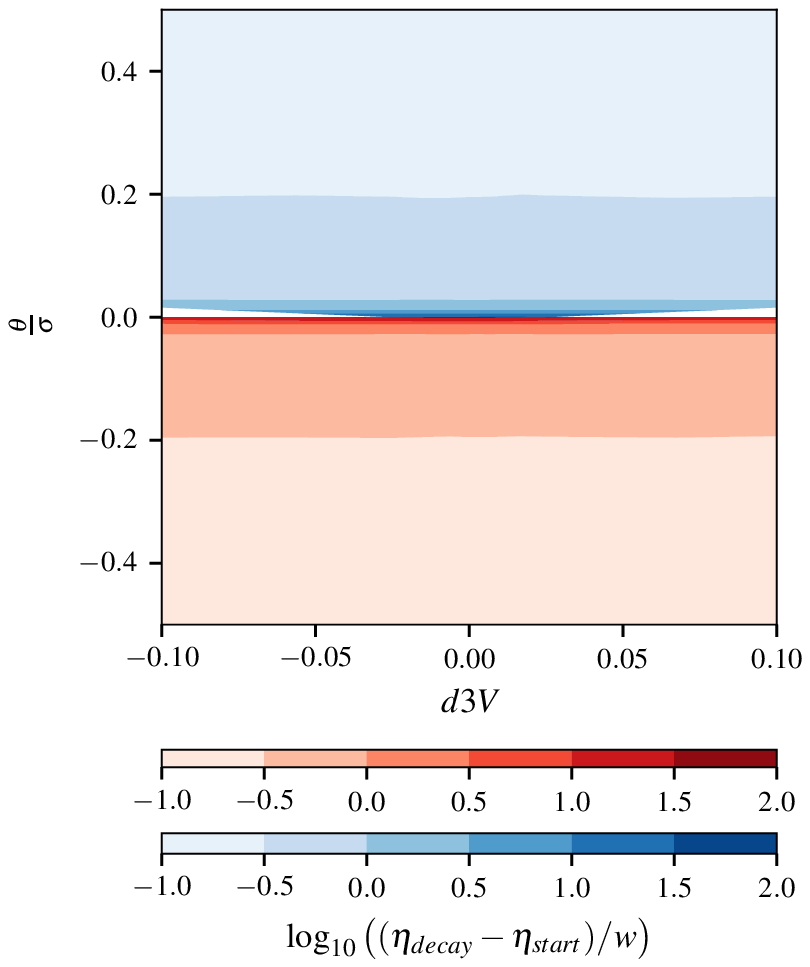}
	\end{subfigure}\hfill%
	\begin{subfigure}[t]{0.5\textwidth}
		\label{fig:d3V_init_mean_deltaV=0,0_init_sigma=0,0625}
		\includegraphics[width=3.375in]{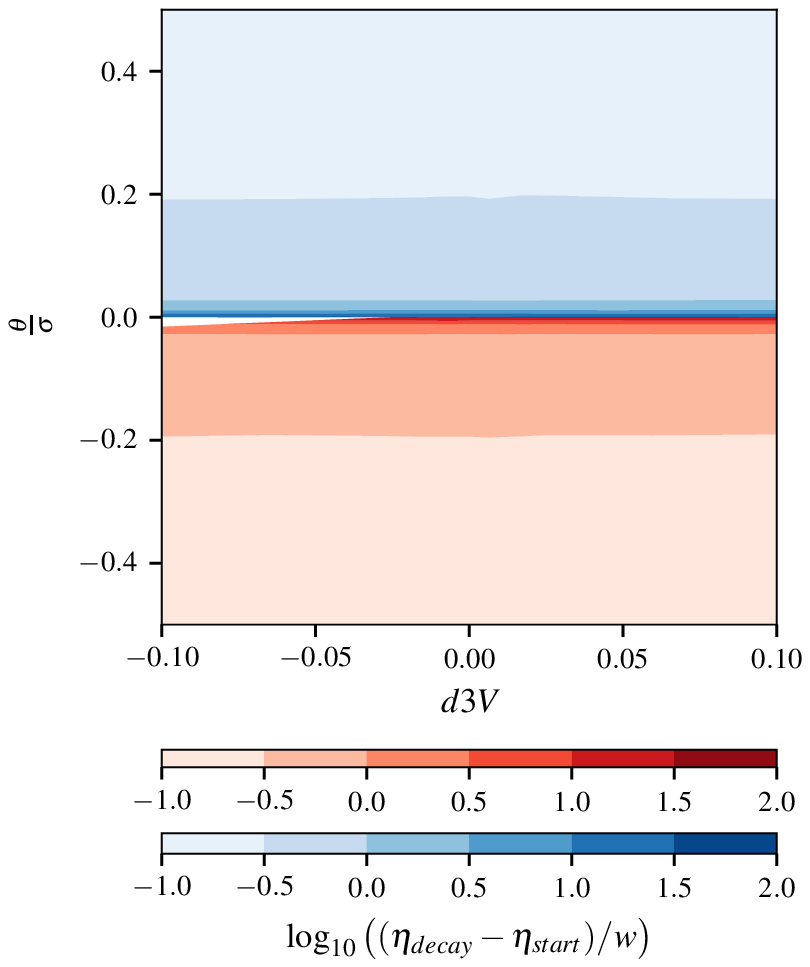}
	\end{subfigure}\hfill%
	\caption{Dependence of the decay time $\eta_{decay} - \eta_{start}$ on $d3V$ and the mean value of initialization distribution $\theta$ with $\sigma = 1$ (left panel) and $\sigma = 0.0625$ (right panel) for models with $\delta V = 0$. \protect\label{fig:decay_time_d3V_init_mean}}
\end{figure*}

Plots from figures \ref{fig:decay_time_deltaV_init_mean} and \ref{fig:decay_time_d3V_init_mean} illustrate influence of bias in the initial distribution. The estimated decay time of networks in function of degeneracy of minima $\delta V$ of a~potential and the mean value $\theta$ of the field at the initialization is plotted in the figure \ref{fig:decay_time_deltaV_init_mean} for two values of the standard deviation of the initialization distribution $\sigma=1$ and $\sigma=0.0625$. It turns out that the shift of initial mean value of the field strength from the position of the local maximum of the potential determines the fate of the network. Even small initial bias toward unstable vacuum makes the network decay into this minimum. Moreover, for both wide and narrow initial distributions the effect is nearly insensitive to values of both $\delta V$ and $d3V$ if the network decay into unstable vacuum, thus details of the shape of the potential do not influence the evolution of the network. On the other hand, if bias is in the direction toward the global minimum of the potential and the initial distribution had large standard deviation $\sigma$, only potentials with nearly degenerate minima lead to long living networks. For initial distributions with small standard deviations this effect is much smaller and decay time of networks depends mainly on the inital value of the field strength $\theta$. Finally, narrow distributions lead to formation of long living networks only when they are very weakly biased.

The figure \ref{fig:decay_time_d3V_init_mean} shows dependence of the decay conformal time on the scale of asymmetry of the potential $d3V$ and the bias of initial distributions $\theta$ for large $\sigma = 1$ and small $\sigma = 0.0625$ standard deviations. For both cases the final state of the evolution is determined by the mean value of the field $\theta$ at the initialization time. Lifetimes of networks are in both cases nearly independent of asymmetry of the potential with only slight increase for nearly symmetric potentials. Formation of long living networks is possible only with small bias of the initial field strength distribution.

Strong dependence on the bias we find is consistent with our earlier studies~\cite{Krajewski:2016vbr,Krajewski:2017czs, Krajewski:2019vix} of the dynamics of domain walls of the Higgs field. We could easily produce networks of domain walls decaying into electroweak vacuum using biased initial distributions even though this vacuum is strongly disfavoured by both difference of values of the effective potential and asymmetry of it around the local maximum. We have shown that only a~small dominance of lattice sites belonging to the basin of attraction of the electroweak vacuum is needed for ending decay of the network in this vacuum for distributions centered around symmetry preserving field strength equal to $0$.

\begin{figure*}[!ht]
	\begin{subfigure}[t]{0.5\textwidth}
		\label{fig:deltaV_init_sigma_d3V=0,0_init_mean=0,0}
		\includegraphics[width=3.375in]{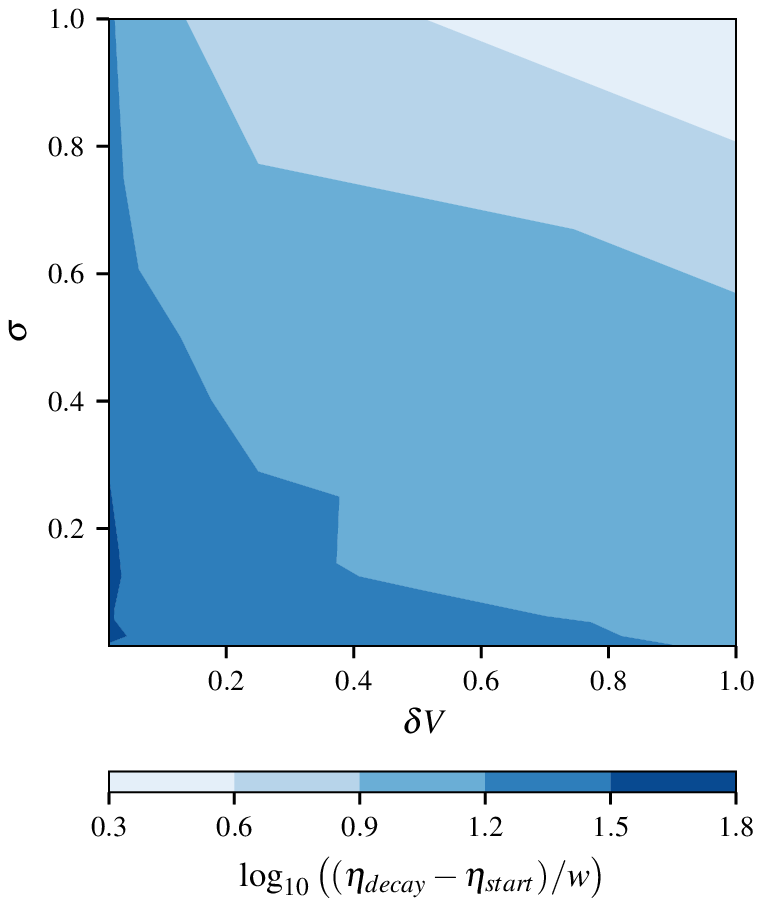}
	\end{subfigure}\hfill%
	\begin{subfigure}[t]{0.5\textwidth}
		\label{fig:deltaV_init_sigma_d3V=0,1_init_mean=0,0}
		\includegraphics[width=3.375in]{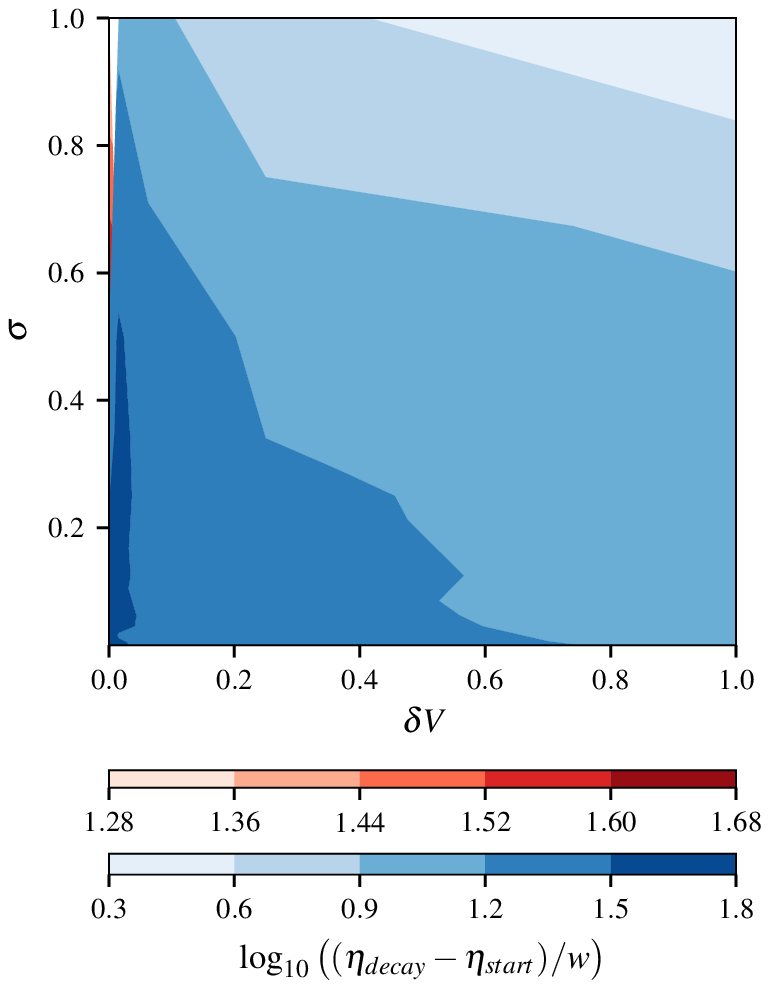}
	\end{subfigure}\hfill%
	\caption{Dependence of the decay time $\eta_{decay} - \eta_{start}$ on $\delta V$ and the standard deviation of initialization distribution $\sigma$ with $\theta=0$ for models with $d3V=0$ (left panel) and $d3V = 0.1$ (right panel). \protect\label{fig:deltaV_init_sigma}}
\end{figure*}

Finally, in the figure \ref{fig:deltaV_init_sigma} we have plotted the extrapolated dependence of the decay time on the level of degeneracy $\delta V$ of the minima and the standard deviation $\sigma$ of the symmetric initial distribution of field strengths for potentials symmetric and highly asymmetric around maximum. Main effect of asymmetry in this case is formation of networks of domain walls that decay into unstable vacuum for models with nearly degenerate minima. As mentioned previously dependence of the decay time of networks decaying into global minimum of the potential on difference of its values at minima $\delta V$ is much stronger than on the asymmetry of the potential around the maximum described by $d3V$, thus both plots in Fig~\ref{fig:deltaV_init_sigma} present similar values. Furthermore, plots in Fig~\ref{fig:deltaV_init_sigma} show that the influence of $\delta V$ on the stability of networks is stronger for initial distributions with large standard deviations~$\sigma$.

\section{Duration of scaling regime\label{scaling_regime}}

In order to deeper understand the issue of metastability of simulated networks we studied appearance of so called scaling regime. Period of simple scaling of average quantities such as surface area or averaged energy of domain walls per Hubble horizon was recognized as an~attractor of the evolution of stable networks of topological defects many years ago. In the scaling regime number of domain walls in Hubble horizon stays nearly constant, sizes of domain and average curvature radius of domain walls are of order of Hubble horizon~\cite{Hindmarsh:1996xv}. In order to maintain these scaling properties domain walls frequently interact with each other, changing their configuration or collapsing into closed walls, to reduce their energy.

It has been found \cite{PhysRevD.39.1558, Lalak:1994qt, Coulson:1995uq,Coulson:1995nv, Larsson:1996sp, Lalak:2007rs,Avelino:2008qy, Correia:2014kqa, Correia:2018tty} that if network of domain walls is unstable due to non-degeneracy of minima of the potential, the evolution in scaling regime ends when the tension of walls no longer compensate pressure produced by the difference of the potential energy density in different vacua on opposite sides of walls and domains of unstable vacuum (the one with vacuum expectation value of the field corresponding to the minima of the potential which is not the global one) collapse, leading to rapid decay of the network. Thus, it is expected that the longer the network will follow the scaling regime the longer it will live.

The velocity depended one scale model (VOS) derived from Nambu-Goto action has a~scaling solution representing this regime for both domain walls and cosmic strings. For former VOS describes a~time evolution of average length scale $L$ which is defined as:
\begin{equation}
L := \frac{\sigma_{wall}}{\rho_{wall}},
\end{equation}
where $\sigma_{wall}$ is a~surface energy density (tension) of domain walls and $\rho_{wall}$ is energy density of walls average over patch of the Universe containing many of them. The second variable in this model is the average velocity norm of walls $v$. Equations of the VOS model take the form:
\begin{equation}
\begin{split}
\frac{dL}{dt} &= (1 + 3 v^2) H L + F(v)\\
\frac{dv}{dt} &= (1 -v^2) \left(\frac{k(v)}{L} - 3Hv\right), \label{eq:VOS}
\end{split}
\end{equation}
where $H$ is value of the Hubble parameter and $k(v)$ is an~effective parameter measuring influence of the Gauss curvature of domain walls on their evolution. Another effective parameter $F(v)$ was introduced in order to account the energy lost of the network. Both $k(v)$ and $F(v)$ as an~effective parameters need to be tuned by other methods (mainly numerical simulations). Independently of forms of $k(v)$ and $F(v)$, the set of equations \eqref{eq:VOS} has a simple scaling solution
\begin{align}
L &= \epsilon t, \qquad v=\textrm{const}, \label{eq:scaling_regime}
\end{align}
for the scale factor with power law dependence on time $a \propto t^\lambda$.

According to \eqref{eq:scaling_regime}, the average energy density of domain walls $\rho_{wall}$ during the scaling regime evolve as
\begin{equation}
\rho_{wall} \propto \frac{\sigma_{wall}}{t}.
\end{equation}
Direct calculation of $\rho_{wall}$ in numerical simulations is complicated, because it requires integration in direction perpendicular to the surface of wall. When domain walls are large comparing to their width, their energy is approximately proportional to their surface area $S_{wall}$ as:
\begin{equation}
\rho_{wall} \approx \frac{\sigma_{wall} S_{wall}}{H^{-3}}.
\end{equation}
Furthermore, in numerical lattice simulation in order to estimate $S_{wall}$ one compute the comoving area of walls averaged over lattice volume:
\begin{equation}
\frac{A}{V} = \frac{a(t) S_{wall}}{H^{-3}} \propto \frac{a(t)}{t}.
\end{equation}
We calculated $A$ according to algorithm presented in \cite{Press:1989yh}.

Finally, in radiation domination epoch $A$ scales as $A \propto t^{-\frac{1}{2}} \propto \eta^{-1}$. Many numerical studies basing on lattice simulations confirmed appearance of such scaling in the evolution of domain walls \cite{Press:1989yh, Lalak:1996db, Garagounis:2002kt, Oliveira:2004he, Avelino:2005pe, Lalak:2007rs,Leite:2011sc, Leite:2012vn, Martins:2016ois}. Moreover, using numerical simulations performed for simple models with degenerated minima of the potential authors of \cite{Hiramatsu:2013qaa} estimated proportionality coefficient
\begin{equation}
\frac{A}{V} = \mathcal{A} \eta^{-1},
\end{equation}
to be of the order of $\mathcal{A} \simeq 0.8 \pm 0.1$.

In order to capture this effect we perform a linear regression and find the longest period such that the evolution followed
\begin{equation}
\log \left( \frac{A}{V} \right) = -\nu \log{\eta} + \log{\mathcal{A}} \protect\label{eq:fit}
\end{equation}
where $\nu$ and $\log{\mathcal{A}}$ were the fitted parameters. In our numerical procedure we dynamically estimated the beginning and the end of the scaling regime. We accepted the period for which the score $R^2$ of linear regression is bigger than $0.8$ with $R^2$ given by:
\begin{equation}
R^2 = \left(1 - \frac{\sum_i( y_i-\hat{y}_i)^2}{\sum_i (y_i-\bar{y})^2}\right),
\end{equation}
where $y_i$ is the value computed in simulations, $\hat{y}_i$ is the predicted value and $\bar{y}=\sum_i y_i$.

\begin{figure*}[!ht]
	\begin{subfigure}[t]{0.5\textwidth}
		\label{fig:d3V=-0.00625,deltaV=0.015625init_mean=0.0init_sigma=0.0125}
		\includegraphics[width=3.375in]{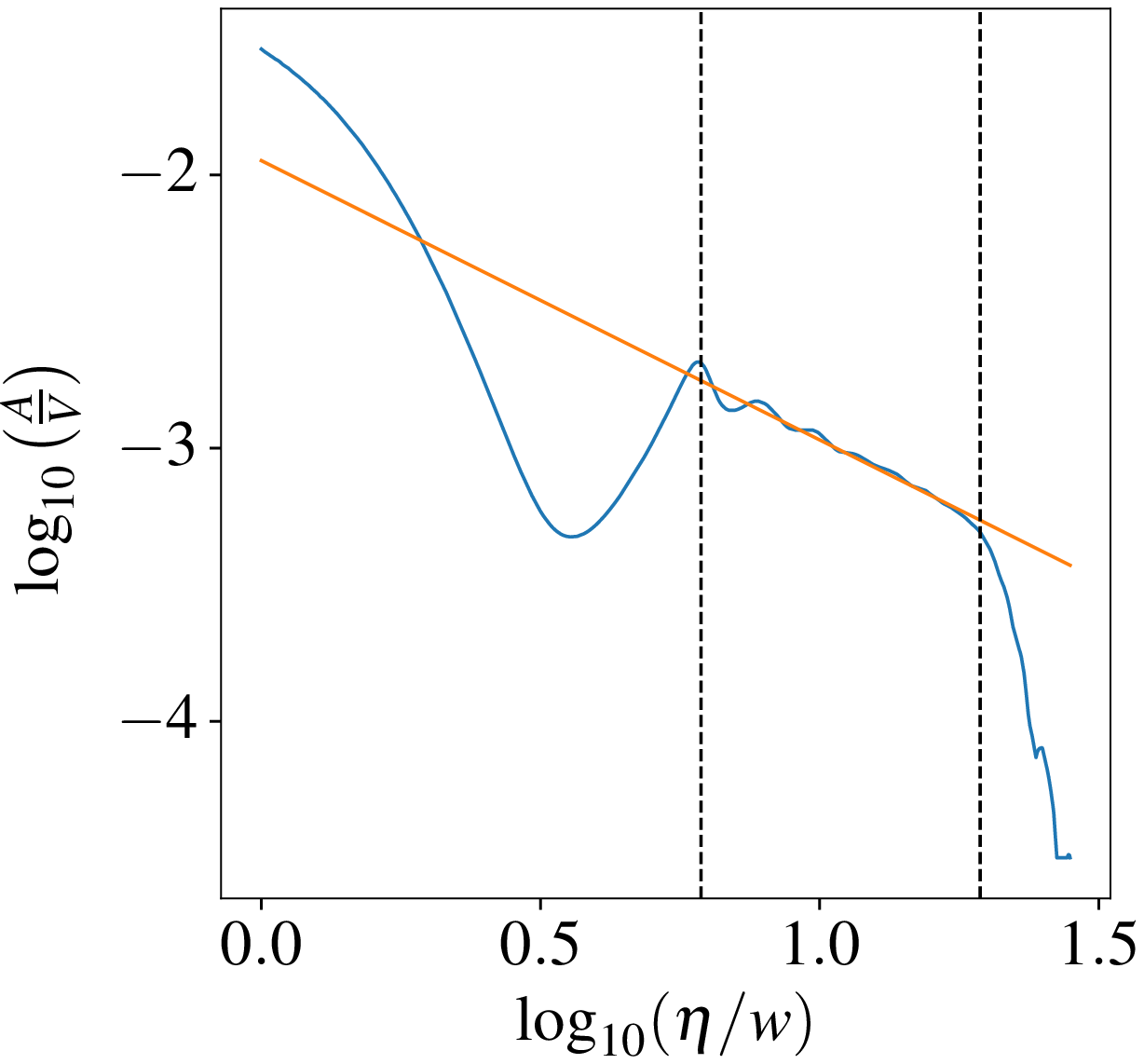}
	\end{subfigure}
	\begin{subfigure}[t]{0.5\textwidth}
		\label{fig:d3V=-0.025,deltaV=0.0init_mean=0.0init_sigma=0.05}
		\includegraphics[width=3.375in]{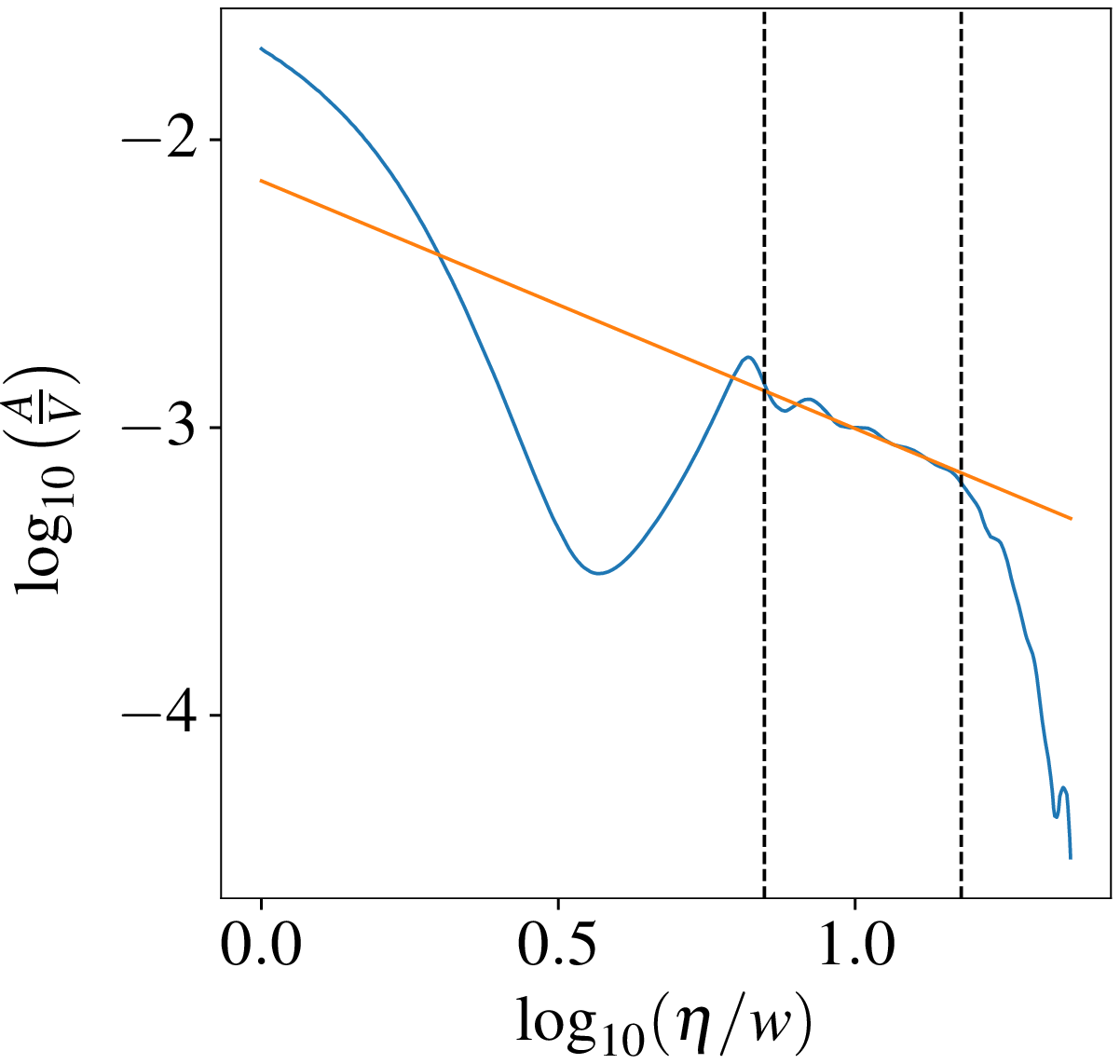}
	\end{subfigure}\hfill%
	\caption{The evolution of conformal surface area of domain walls per unit volume $\frac{A}{V}$ in function of conformal time $\eta$ (blue) and the fitted scaling behavior defined by eq. \eqref{eq:fit} (orange) for the best (left panel) and the worst (right panel) fits obtained by procedure described in the main text. Vertical dashed lines correspond to the estimated beginning and end of the scaling regime. \protect\label{fig:scaling_examples}}
\end{figure*}

Plot on the left panel of the figure \ref{fig:scaling_examples} shows the best obtained fit ($R^2$ closest to $1$) and the one on the right panel the worst fit that we accepted. The visual inspection of these plots reveals that the main contribution to our quality measure $R^2$ come from edges of the fitted period. In our procedure the end of scaling regime is effectively determined by the deviation from linear relationship of eq. \eqref{eq:fit} saturating our bound on $R^2$. On the other hand, we do not want to restrict $R^2$ to be too close to $1$, because we believe that early period when oscillations of surface area appear (as visible in the figure \ref{fig:scaling_examples}) should be included as a~part of the scaling regime.

Obtained fitted exponent $\nu$ of scaling behavior ranges from $\nu = 0.81$ up to $\nu = 1.0$. The highest obtained value of $\nu$ is in good agreement with predictions of semi-analytical VOS model while the lowest one correspond to the network decaying a~bit slower than one expects from thin walls approximation. However, many authors \cite{Press:1989yh, Lalak:1996db, Lalak:1994qt} have noted that numerical simulations performed in the past predict $\nu$ to be lower than $1$ and even as low as $\nu = 0.6$ mentioned in \cite{Lalak:1996db}. Scaling parameter $\mathcal{A}$ obtained by this procedure ranges from $0.08$ up to $0.34$. These values are smaller than the one calculated previously by authors of \cite{Hiramatsu:2013qaa}. The probable cause of discrepancy is asymmetry of the potential in our simulations which destabilize domain walls in contrast to stable domain walls in model with the symmetric potential studied in \cite{Hiramatsu:2013qaa}. The duration of the period to which we fitted scaling law from eq.~\eqref{eq:fit} ranges from $8.5\, w$ up to $29.7\, w$ in units of the width of walls $w$.

Described procedure performed on the data gathered from lattice simulations revealed that only networks formed from unbiased or weakly biased initial distributions enter scaling regime. Moreover, only small values of $\delta V$ parameter are allowed, thus the approximated degeneracy of minima is needed to observe scaling behavior of the network. We have not observed networks evolving in scaling regime in models with difference of values of the potential in minima larger than $\delta V = 0.0625$. Moreover, we have found that networks formed from narrow initial distributions enter scaling regime easier with smaller fluctuations of the conformal surface area. In case of wide distributions with standard deviation $\sigma = 1$ oscillations were so large that we were not able to reliably determine period of the scaling regime.

\section{Gravitational waves emitted from domain walls\label{spectrum}}
After the recent discovery of gravitational waves in collaborating LIGO and Virgo \cite{Abbott:2016blz} experiments the direct detection of primordial gravitational waves emitted from cosmological sources is widely discussed. Gravitational waves are a new unique source of information about the early Universe. After emission they interact very weakly with other constituents of the Universe and simply continue to propagate till present time. Thus, they still carry direct information about processes which produced them.

Topological defects are one of the possible sources of GWs. During the decay of the network, energy density stored in defects is transferred to other degrees of freedom including GWs. In order to determine if GWs produced by domain walls can be observed in current or future detectors one needs to estimate their strength and frequencies. In case of possible detection its distinction form signal produced by other sources will relay on our knowledge of the shape of the spectrum.

However, direct calculation of the spectrum of gravitational waves in lattice simulations encounters many complications. The algorithm of PRS~\cite{Press:1989yh} cannot be used, because the modification of the equation of motion disturbs the dynamics of the short wavelength fluctuations. For the unmodified eom the width of domain walls decreases as $\propto a^{-2}$. This effect significantly restricts the dynamical range of the simulation. Moreover, as noted in~\cite{Hiramatsu:2013qaa}, the algorithm presented there which is widely used produces a~spectrum that diverges as $k^3$ for random initialization of the field strength. Insufficient number of small wave-vectors fitting into finite lattice is another problem. Our experience shows that currently available methods are not precise enough to fully track subtle effect of asymmetry of the potential on the shape of the spectrum of GWs.

Thus, in this paper we limit our research to estimation of influence of asymmetry of the potential on the peak frequency and the amplitude of the spectrum of GWs basing on the semi-analytical approximation. We postpone direct calculation of the spectrum for future work. With that said, we expect that the complete spectra could carry interesting information on the main reason for the rapid decay of the network. Features of the potential have been shown to influence the GW spectra produced by bubble collisions in first order phase transitions~\cite{Cutting:2020nla} which is a related system. In fact, the GW spectra we show here are identical to the ones produced with the envelope approximation~\cite{Kosowsky:1992vn} first envisaged as an approximate description of bubble collisions taking place in a first order phase transition.

We are interested in the spectrum of GWs' energy density $\rho_{GW}$ per unit logarithmic frequency interval as a~fraction of the critical density $\rho_c$:
\begin{equation}
\Omega_{GW} (\eta) := \frac{1}{\rho_c(\eta)}\frac{d \rho_{GW}}{d \log |k|} (\eta,k).
\end{equation}
During scaling regime domain walls' averaged energy density decreases with expansion slower than energy density of the radiation, especially gravitational waves. Thus, one expects that the peak of the spectrum of the waves emitted from domain walls will be located at the frequency corresponding to the Hubble scale around the time of the decay of the network~\cite{Hiramatsu:2013qaa}. For nearly degenerate minima one can estimate the energy density of GWs at the peak using semi-analytic expression \cite{Hiramatsu:2013qaa,Kitajima:2015nla}:
\begin{equation}
\left. \Omega_{GW} (\eta_{dec}) \right|_{peak} = \frac{\tilde{\epsilon}_{GW} \mathcal{A}^2 {\sigma_{wall}}^2}{24 \pi {H_{dec}}^2 {M_{Pl}}^4}, \label{eq:peak_decay}
\end{equation}
where $\tilde{\epsilon}_{GW}$ is the efficiency parameter determined in numerical simulations of $\lambda \phi^4$ model \cite{Kitajima:2015nla} to be equal to $\tilde{\epsilon}_{GW} \simeq 0.7$. In our computations we assume that $\tilde{\epsilon}_{GW}$ in considered models does not differ much from the value computed previous and should be of order $\BigO{1}$.

Tension $\sigma_{wall}$ (energy density per unit surface area) of domain walls for considered models calculated according to algorithm presented in the subsection \ref{width} manifests weak dependence on parameters $\delta V$ and $d3V$ of our family of potentials. For models in question $\sigma_{wall}$ changes slightly, ranging from $51.0\, w^{-3}$ up to $102.3\, w^{-3}$.

Peak amplitude given by \eqref{eq:peak_decay} corresponds to the value at the time when the network decays. Red-shifting the value up to today we find~\cite{Kamionkowski:1993fg}
\begin{equation}
\begin{split}
\Omega_{GW} (\eta_{0})&=\left(\frac{a(\eta_{dec})}{a(\eta_{0})}\right)^4\left(\frac{H(\eta_{dec})}{H(\eta_{0})}\right)^4 \Omega_{GW} (\eta_{dec})\\
&=1.67\times 10^{-5} h^{-2}\left(\frac{100}{g_*(\eta_{dec})}\right)^\frac13 \Omega_{GW} (\eta_{dec}).
\end{split}
\end{equation}
Using this to rewrite~\eqref{eq:peak_decay} the amplitude of the peak of the GW spectrum measured today can be estimated as:
\begin{equation}
\begin{split}
\left. \Omega_{GW} (\eta_{0}) \right|_{peak} &= 4.6\times 10^{-81}
\mathcal{A}^2 \left(\frac{\rm GeV}{H_{dec}}\right)^2 \\
&\times \left(\frac{\sigma_{wall}}{
\rm GeV^3}\right)^2 h^{-2}\left(\frac{100}{g_*(\eta_{dec})}\right)^\frac13.
\label{eq:today_Omega}
\end{split}
\end{equation}

In addition, we have to calculate present day frequency of the peak. The wavelength $\lambda(\eta)$ of the GW with the comoving wave vector $k$ at the conformal time $\eta$ satisfies:
\begin{equation}
k a(\eta)^{-1} \lambda(\eta)=2 \pi. \label{eq:wavelength}
\end{equation}
Equating $\frac{k}{2\pi}$ from  eq. \eqref{eq:wavelength} for the time of the decay $\eta_{dec}$ and the present time $\eta_0$ we estimated the red-shift of the wave frequency to be equal to:
\begin{equation}
\left. f_{0}\right|_{peak} = \frac{a(\eta_{dec})}{a(\eta_0)} H_{dec} = 1.63\times 10^2 \left(\frac{H_{dec}}{\rm GeV}\right)^\frac{1}{2}\; \textrm{Hz}, \label{eq:redshifted_f}
\end{equation}
where we assume that wavelength of the peak is equal to Hubble radius $\lambda_{dec} = {H_{dec}}^{-1}$. We performed our simulations assuming that the evolution of domain walls took place during radiation domination era, thus the lowest value of $H_{dec}$ for which our numerical results are reliable corresponds to matter-radiation equality $H_{EQ}$ which gives the lower bound on the frequency of the peak to be of the order of $\left. f_{0}\right|_{peak} \gtrsim 10^{-16}\; \textrm{Hz}$.

Finally, it is convenient to express both eqs. \eqref{eq:today_Omega} and \eqref{eq:redshifted_f} in terms of the lifetime of the network $\eta_{dec}$ and the width of domain walls $w$:
\begin{align}
\left. \Omega_{GW} (\eta_{0}) \right|_{peak} =& 0.29 \times 10^{-77} \mathcal{A}^2 \nonumber \\
&\times \left(\frac{\eta_{dec}}{w}\right)^4 \left(\frac{\sigma_{wall}}{w^{-3}}\right)^2 \left(\frac{
\rm GeV^{-1}}{w}\right)^4, \label{eq:Omega_simulation} \\
\left. f_{0}\right|_{peak} =& 3.3 \times 10^1  \left(\frac{w}{\eta_{dec}}\right) \left(\frac{
\rm GeV^{-1}}{w}\right)^\frac{1}{2}\; \textrm{Hz}, \label{eq:f_simulation}
\end{align}
where we have assumed the scale factor dependence on conformal time given by the eq. \eqref{eq:scale_factor} with $\eta_{start} = c\, l = 0.04\, w$, as in our lattice simulations. Thus, we see that both the peak frequency and amplitude decrease as the width of domain walls increases. On the other hand, with increasing lifetime of the network the amplitude of the peak increases and the frequency decreases.

\begin{figure*}[ht]
\centering
	\includegraphics[width=0.75\textwidth]{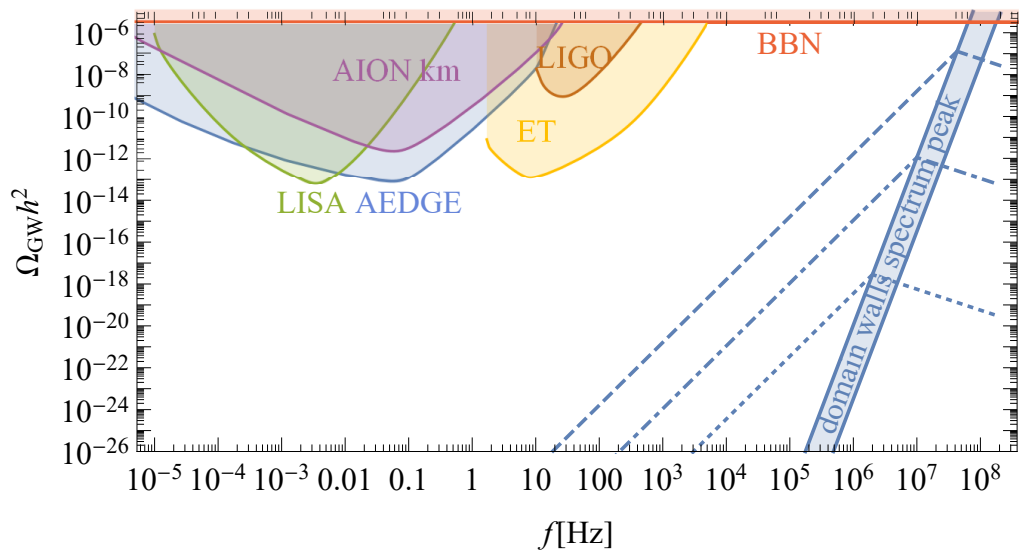}
	\captionsetup[sub]{labelformat=empty,justification=raggedleft,singlelinecheck=false}
	\caption{The blue band shows hypothetical peak amplitudes of GWs emitted from cosmological domain walls as a function of the peak frequency $f$. The width of the band comes from the possible range on the prefactor controlling the amplitude of the signal (See Eq.~\ref{eq:GWmax} and Eq.~\ref{eq:GWmin}). The shape of the spectra peaking in the allowed region is indicated by the dashed blue lines. These should be compared to predicted sensitivities of currently operating and planned detectors LIGO~\cite{Thrane:2013oya,TheLIGOScientific:2014jea,TheLIGOScientific:2016wyq,LIGOScientific:2019vic}, LISA~\cite{Bartolo:2016ami,Caprini:2019pxz}, AEDGE~\cite{Bertoldi:2019tck}, AION-1km~\cite{Badurina:2019hst}, ET~\cite{Punturo:2010zz,Hild:2010id} as well as upper bound induced by the CMB/BBN~\cite{Smith:2006nka,Henrot-Versille:2014jua}. \protect\label{fig:sensitivites}
	}
\end{figure*}

We have estimated overall factors present in eqs. \eqref{eq:Omega_simulation} and \eqref{eq:Omega_simulation} basing on values of $\mathcal{A}$, $\eta_{dec}$ obtained in simulations in which networks entered scaling regime and previously computed $\sigma_{wall}$. The maximal value of the prefactor in eq. \eqref{eq:Omega_simulation} obtained in this way is equal to:
\begin{align} \label{eq:GWmax}
\left. \Omega_{GW}^{max} (\eta_{0}) \right|_{peak} &= 0.1 \times 10^{-66} \left(\frac{1 \frac{\hbar c}{\textrm{GeV}}}{w}\right)^4,\\
 \left. f_{0}^{max} \right|_{peak} &= 0.7 \left(\frac{1 \frac{\hbar c}{\textrm{GeV}}}{w}\right)^\frac{1}{2}\; \textrm{Hz},
\end{align}
where the frequency of the peak for this network is denoted as $f_{0}^{max}$. On the other hand, the minimal prefactor computed from data from simulations is equal to:
\begin{align} \label{eq:GWmin}
\left. \Omega_{GW}^{min} (\eta_{0}) \right|_{peak} &= 0.6 \times 10^{-68} \left(\frac{1 \frac{\hbar c}{\textrm{GeV}}}{w}\right)^4,\\
\left. f_{0}^{min} \right|_{peak} &= 1.3 \left(\frac{1 \frac{\hbar c}{\textrm{GeV}}}{w}\right)^\frac{1}{2}\; \textrm{Hz}.
\end{align}
These predictions for the peak amplitude and its peak frequency are shown in the figure \ref{fig:sensitivites} together with sensitivities of current and proposed detectors of GWs. The main difference between our results and the typical assumption of a scaling network comes from the impact of the short lifetime $\eta_{dec}$ in eq.~\eqref{eq:f_simulation}. It is common in the literature to assume the network decays just before dominating the expansion which gives the largest possible abundance and much lower peak frequency. In our results we can see instead that only a very energetic network is capable of producing a strong signal. This very energetic network, however, has to be created at an appropriately high energy scale and the corresponding peak frequency is also very high.

Detection of GWs emitted from domain walls in models with asymmetric potentials is a daunting task. Frequencies below $1\; \textrm{kHz}$ in which interferometer based detectors are sensitive corresponds to domain walls with energy below $10^{6}\; \textrm{GeV}$ which produce an extremely weak signal. Stronger signal can come from domain walls at a much higher energy scale for example the GUT scale $\sim10^{15}\; \textrm{GeV}$. However, then the signal is characterized by much higher peak frequency around $1\; \textrm{MHz}$--$1\; \textrm{GHz}$.
Even though, certain mechanisms for detection GWs at very high frequencies were proposed~\cite{Aggarwal:2020olq}, their predicted reach in terms of abundance is still above the current lower bounds coming from CMB and BBN observations making detection of GWs produced by domain walls extremely difficult.

\section{Conclusions\label{summary}}
In this paper we investigated dependence of stability of cosmological domain wall networks on the shape of the potential and the initial probability distribution of the field strength. Our main aim was to determine how the following four factors influence the evolution of the networks:
\begin{itemize}
	\item difference of values of the potential at minima,
	\item asymmetry of potential around local maximum separating minima,
	\item width of the initial distribution of the field strength,
	\item bias (shift with respect to position of the local maximum) of the initial distribution.
\end{itemize}
We identified relative importance of these factors on lifetime of the networks. Effects of some of these factors were studied in the past \cite{Larsson:1996sp, PhysRevD.65.025002, Lalak:2007rs, Correia:2014kqa, Correia:2018tty}, however broad analysis was performed for the first time. Moreover, the shape of the potential around local maximum was mentioned as a factor that can influence the stability of the network \cite{Lalak:2007rs}, but this hypothesis was not verified till now. 

In order to study the influence of the shape of the potential on the dynamics of the network we extended the typically used toy quartic potential. We constructed a family of potentials whose shape around the potential barrier and the level of degeneracy of minima can be set independently. We parameterized these features as the value of the third derivative at the local maximum separating minima denoted as $d3V$ and the difference of values of the potential at minima denoted as $\delta V$. The width of walls which determines the energy scale of the problem is constant in this family, thus the evolution of networks for all the potentials in the family can be directly compared.

We studied the evolution of networks of domain walls in models given by potentials from the constructed family using lattice simulations based on the constant width PRS algorithm. After preforming thousands of simulations we were able to determine how the above mentioned factors influence the evolution of the networks. Results of our simulations allowed us to estimate the relative importance of these factors.

We found that the final state of the decay of the network is determined by the bias of the initial probability distribution. Even though, other factors can shorten or enlarge the life-time of the network, the excess of lattice points belonging to one of basins of attraction of minima of the potential drive the evolution of domain walls into corresponding vacuum.

When the initial distribution is symmetric with respect to the position of the local maximum, the fate of the network is determined mainly by the difference of values of the potential in the minima. We observed decay into vacuum corresponding to the minimum with higher value of the potential, only for potentials with nearly degenerated minima. As one may expect, asymmetry of the potential around local maximum which pushed the evolution of these networks toward unstable vacuum has stronger effect for narrower, more condensed around the local maximum, initial distributions.
On the other hand, when minima of the potential are nearly degenerate, asymmetry around its local maximum toward higher energy minimum may stabilize networks decaying into this vacuum for weakly biased initial distributions.

Even though, bias of initial distribution, difference of values of the potential at minima and asymmetry of potential around local maximum are listed together in the literature as factors affecting stability of networks, they are not equally important. Our numerical simulations prove that a hierarchy of strength of influence produced by these factors does exist. In the past the importance of the asymmetry of initial distribution was recognized. However, the strength of its influence on the stability of domain walls in comparison to the difference of the values of potential in minima was underestimated, since nearly symmetric potentials were considered in early studies. We have shown that initial bias can trigger decay into local minimum of the potential, even when it is highly disfavored by the shape of the potential. Moreover, the details of the shape turned out to affect stability of networks weaker that expected and it is the energy difference between the minima that determines the leading effect.

In order to better understand the issue of the metastability of networks of domain walls we extended our studies by searching for sings of so called scaling regime which was recognized as an~attractor solution of the evolution of network of topological defects early in the history of studies of these objects~\cite{Vilenkin:2000jqa}. It is characterized by the simple scaling of averaged statistical variables such as volume average of the surface area of domain walls with respect to expansion of the Universe. During this regime domain walls interact frequently with each other preserving the scaling behavior, leaving nearly constant number of walls in each Hubble horizon. Long living networks are expected to enter this regime. Metastable networks may stay in it for a~very long time, till they rapidly, completely decay.

Our numerical procedure for finding the simple power-law scaling of the conformal surface area of walls in function of the conformal time has revealed that only a~small fraction of simulated networks entered the scaling regime. We have found that the bias of the initial probability distribution of the field strength prevents scaling. Moreover, only models with potentials with nearly degenerated minima allow the evolution of network in the scaling regime.

We have computed the exponent $\nu$ of the power law describing the evolution of networks in the scaling regime. For performed simulations it ranges from $\nu = 0.8$ up to $\nu = 1$. Obtained values are consistent with those mentioned in the literature \cite{Press:1989yh, Lalak:1994qt, Larsson:1996sp, Lalak:1996db}. Moreover, we have determined the scaling parameter $\mathcal{A}$ describing scaling of the averaged domain walls' energy density. From data gathered in our simulations we estimated $\mathcal{A}$ to be in the range $0.08-0.34$. these values are smaller, than the one obtained by the authors of \cite{Hiramatsu:2013qaa}. The possible source of discrepancy is the instability of domain walls in models studied in this paper. In~\cite{Hiramatsu:2013qaa} stable domain walls in model with exactly symmetric potentials and initial distributions were simulated.

Using semi-analytical expressions for amplitude and location of the peak in the spectrum of gravitational waves (GWs) emitted from domain walls we have estimated these quantities using $\mathcal{A}$ computed from data gathered in our numerical simulations. We have found that domain walls in models with asymmetric potentials would produce extremely weak signals at frequencies below $1\; \textrm{kHz}$ which is an upper bound on the sensitivity of current and planned interferometer detectors.

It is well known that in order to produce signal observable in currently running detectors domain walls have to be metastable with long decay time of the order $\gtrsim 10^6 \textrm{--} 10^7 \frac{w}{c}$ in the units of walls' width $w$ \cite{Hiramatsu:2013qaa}. As we have shown asymmetry of the potential destabilizes the networks of domain walls, thus such long lifetime is excluded when asymmetry of the potential is not fine-tuned to be extremely small.

Producing long living networks of domain walls may seem not too problematic because symmetry of the potential can be naturally protected by symmetry of the model. However, it was realised in the past \cite{Lalak:1994qt, Lalak:1996db, PhysRevD.65.025002, Lalak:2007rs} (and we have confirmed this observation) that the bias of the initial distribution of the field strength toward one of minima of the potential destabilize the network. Forcing the initial distribution of the field to be centered at the local maximum separating the minima is far less obvious and depends on the processes responsible for fluctuations of the field.

For example, it is well known that inflation produces superhorizon fluctuations which are nearly Gaussian. However, the mean value of the distribution is not affected by the process and is determined by the preinflationary evolution of the field. Thus, producing the metastable network of domain walls living long enough to produce GWs signal in sensitivity range for interferometer detectors can be much more unnatural than is commonly believed.

On the other hand, even short living networks can produce a GWs signal with appreciable abundance if their energy scale is high enough (or in other words width $w$ small enough).
However, then the frequency of the peak of the spectrum is $10^7$ Hz or more, well above the sensitivity range of current and planned detectors making detection of GWs produced by domain walls at very high frequencies also a very difficult prospect.

\begin{acknowledgments}
	This work has been supported by the Polish National Science Center grants 2019/32/C/ST2/00248, 2018/29/N/ST2/01743, 2018/31/D/ST2/02048 and 2017/27/B/ST2/02531.
	The project is co-financed by the Polish National Agency for Academic Exchange within Polish Returns Programme under agreement PPN/PPO/2020/1/00013/U/00001. This research was supported in part by PL--Grid Infrastructure. J.H.K. acknowledges hospitality of CP3-Origins, where parts of this work has been done.
\end{acknowledgments}

\appendix

\section{Analytical solution of domain wall profile and its width\label{app:toy_model}}
In this appendix we will describe our general setup in a~simple model \eqref{symmetric_lagrangian_density}. The eom derived from \eqref{symmetric_lagrangian_density} in the Minkowski gravitational background takes the form:
\begin{equation}
\frac{\partial^2 \phi}{{\partial t}^2} -\Delta \phi = - \frac{\partial V}{\partial \phi} = 4 V_0 \left(\frac{\phi^2}{{\phi_0}^2}-1\right)\frac{\phi}{{\phi_0}^2}. \label{eq:toy_eom}
\end{equation}

We are interested in a time independent solution (soliton solution). We will consider planar walls i.e. solutions with the translational symmetry in two space dimensions. Assuming \mbox{$\phi(t,x,y,z)=\varphi(x)$}, our Lagrangian density \eqref{symmetric_lagrangian_density} simplifies to
\begin{equation}
\mathcal{L}=-\frac{1}{2} \varphi'^2 - V\left(\varphi\right)=-\frac{1}{2} \varphi'^2- V_0 \left(\frac{\varphi^2}{{\phi_0}^2}-1\right)^2, \label{eq:toy_lagrangian_density_simplified}
\end{equation}
where prime is a derivative with respect to $x$.
This Lagrangian density has the translational symmetry in $x$ and the corresponding conservation law. The associated conserved quantity is
\begin{equation}
E=\frac{1}{2} \varphi'^2 - V\left(\varphi\right)=\frac{1}{2} \varphi'^2- V_0 \left(\frac{\varphi^2}{{\phi_0}^2}-1\right)^2. \label{eq:energy}
\end{equation}
Using conservation of $E$ we get first-order differential equation:
\begin{equation}
\begin{split}
\varphi' &= \pm \sqrt{2\left(E+V\left(\varphi\right)\right)}\\
&=\pm \sqrt{2\left(E+V_0 \left(\frac{\varphi^2}{{\phi_0}^2}-1\right)^2\right)},
\end{split}
\end{equation}
which can be easily integrated,
\begin{equation}
\begin{split}
x(\varphi_2)-x(\varphi_1)&= \pm \int_{\varphi_1}^{\varphi_2} \frac{d\varphi}{\sqrt{2\left(E+V\left(\varphi\right)\right)}} \\
&=\pm \int_{\varphi_1}^{\varphi_2} \frac{d\varphi}{\sqrt{2\left(E+V_0 \left(\frac{\varphi^2}{{\phi_0}^2}-1\right)^2\right)}}, \label{integral}
\end{split}
\end{equation}
for appropriate values of $\phi_1$ and $\phi_2$. Choosing $x_1=\phi_1=0$ we get our soliton solution
\begin{equation}
\varphi(x) = \phi_0 \tanh \left(\frac{\sqrt{2 V_0}}{\phi_0}x\right) = \phi_0 \tanh \left(\frac{\pi x}{w_0}\right), \label{eq:soliton}
\end{equation}
where $w_0=\frac{\pi \phi_0}{\sqrt{2 V_0}}$ is a~width of the wall.
We can also calculate surface potential energy, using
\begin{equation}
\begin{split}
\sigma(x_1,x_2)&:=\int_{x_1}^{x_2} V(\varphi(x)) dx \\
&= \int_{\varphi(x_1)}^{\varphi(x_2)}\frac{V(\varphi) d\varphi}{\sqrt{2\left(E+V\left(\varphi\right)\right)}}. \label{eq:tension}
\end{split}
\end{equation}
Most of the energy of the wall is concentrated at distances of the order of $w_0$ from the center of the wall,
\begin{equation}
\frac{\sigma(-\frac{w_0}{2},\frac{w_0}{2})}{\sigma(-\infty,\infty)} \approx 97 \%.
\end{equation}
This justifies the estimation of the domain wall thickness by the quantity $w_0$.

\section{Parameters of model of asymmetric potentials\label{app:parameters}}
Table \ref{tb:parameters} contains parameters $b$, $d$ and $V_0$ of potentials of the form given by the eq. \eqref{eq:asymmetric_potential} determined as a~solution of the set of eqs. \eqref{eq:final_set}. Potentials with parameters specified by the table \ref{tb:parameters} were used in our lattice simulations in order to model generic asymmetric potentials.
\begin{table}
	\centering
	\begin{tabular}{|c|c|c|c|c|}
		\hline
		$d3V$ & $\delta V$ & $b$ & $d$ & $V_0$ \\ \hline
		-0.00625 & 0.015625 & 1.04147 & 0.016248 & 0.40925 \\
		0.00625 & 0.015625 & 1.04634 & 0.025830 & 0.40663 \\
		-0.00625 & 0.0625 & 1.17135 & 0.069358 & 0.32248 \\
		0.00625 & 0.0625 & 1.17694 & 0.079796 & 0.32011 \\
		-0.00625 & 0.25 & 1.53108 & 0.17416 & 0.22488 \\
		0.00625 & 0.25 & 1.53817 & 0.18552 & 0.22294 \\
		-0.00625 & 0. & 0.997988 & -0.0040296 & 0.51818 \\
		-0.00625 & 0. & 0.997988 & -0.0040296 & 0.51818 \\
		-0.00625 & 1. & 2.17968 & 0.28832 & 0.14973 \\
		0.00625 & 1. & 2.18968 & 0.30110 & 0.14819 \\
		-0.025 & 0.015625 & 1.03424 & 0.0019283 & 0.41318 \\
		0.025 & 0.015625 & 1.05372 & 0.040261 & 0.40269 \\
		-0.025 & 0.0625 & 1.16309 & 0.053830 & 0.32602 \\
		0.025 & 0.0625 & 1.18545 & 0.095588 & 0.31654 \\
		-0.025 & 0.25 & 1.52066 & 0.15739 & 0.22777 \\
		0.025 & 0.25 & 1.54904 & 0.20285 & 0.21999 \\
		-0.025 & 0. & 0.991992 & -0.016099 & 0.52237 \\
		-0.025 & 0. & 0.991992 & -0.016099 & 0.52237 \\
		-0.025 & 1. & 2.16513 & 0.26964 & 0.15200 \\
		0.025 & 1. & 2.20516 & 0.32082 & 0.14585 \\
		-0.1 & 0.015625 & 1.00623 & -0.054847 & 0.42873 \\
		0.1 & 0.015625 & 1.08424 & 0.098838 & 0.38681 \\
		-0.1 & 0.0625 & 1.13134 & -0.0069446 & 0.33999 \\
		0.1 & 0.0625 & 1.221 & 0.16062 & 0.30207 \\
		-0.1 & 0.25 & 1.48133 & 0.093279 & 0.23905 \\
		0.1 & 0.25 & 1.59554 & 0.27625 & 0.20786 \\
		-0.1 & 0. & 0.968597 & -0.064163 & 0.53900 \\
		-0.1 & 0. & 0.968597 & -0.064163 & 0.53900 \\
		-0.1 & 1. & 2.11155 & 0.20006 & 0.16068 \\
		0.1 & 1. & 2.27404 & 0.40755 & 0.13590 \\
		0. & 0.015625 & 1.0439 & 0.021035 & 0.40794 \\
		0. & 0.0625 & 1.1741 & 0.074568 & 0.32129 \\
		0. & 0.25 & 1.5346 & 0.17982 & 0.22391 \\
		0. & 1. & 2.1847 & 0.29467 & 0.14896 \\
		\hline
	\end{tabular}
	\caption{Numerical solutions to the set of equations \protect\eqref{eq:final_set}. \protect\label{tb:parameters}}
\end{table}

\bibliography{SDWMAP}
\end{document}